\documentclass[twocolumn,english,aps,prd,reprint,floatfix,notitlepage,footinbib,preprintnumbers,superscriptaddress,altaffilletter]{revtex4-2}

\usepackage{amsmath,amssymb,amsfonts}
\usepackage{hyperref,breakurl,cleveref,url}
\usepackage{color}

\usepackage{graphicx}
\usepackage[export]{adjustbox}

\usepackage{accents,mathrsfs,mathtools,dsfont}

\renewcommand{\paragraph}[1]{%
    \textit{#1}.---%
}

\def\skip{\vskip1.5pt}
\newcommand\trick[1]{}

\usepackage{enumitem}
\setlist[enumerate]{
    label={},
    leftmargin=2em,
    itemsep=2pt,
    topsep= 2pt,
    partopsep=0pt,
    parsep=0pt,
}

\usepackage{hyphenat}
\let\oldeqref\eqref
\renewcommand{\eqref}[1]{Eq.\,\smash{\oldeqref{#1}}}
\newcommand{\eqrefs}[2]{Eqs.\,\smash{\oldeqref{#1}} and \smash{\oldeqref{#2}}}

\newcommand{\rcite}[1]{Ref.\,\smash{\cite{#1}}}
\newcommand{\rrcite}[1]{Refs.\,\smash{\cite{#1}}}

\newcommand{\fref}[1]{Fig.\,\ref{#1}}

\DeclareMathOperator{\sinc}{sinc}

\DeclareMathOperator{\Exp}{Exp}

\def\mem{\hspace{0.1em}}
\def\hem{\hspace{0.05em}}
\def\nem{\hspace{-0.1em}}
\def\hnem{\hspace{-0.05em}}
\def\hhem{\hspace{0.025em}}
\def\hhnem{\hspace{-0.025em}}

\def\a{\alpha}
\def\b{\beta}
\def\c{{\gamma}}

\def\e{\epsilon}

\def\m{\mu}
\def\n{\nu}
\def\r{\rho}
\def\s{\sigma}

\def\t{\tau}

\def\L{\Lambda}

\def\bgamma{\bar{\gamma}}
\def\bpsi{\bar{\psi}}

\def\tpsi{\tilde{\psi}}

\def\bR{\bar{R}}

\def\mplus{{\mem+\mem}}
\def\mminus{{\mem-\mem}}
\def\mtimes{{\mem\times\hem}}

\def\mtensor{{\mem\otimes\mem}}
\def\moplus{{\mem\oplus\mem}}
\def\mwedge{{\mem\wedge\mem\hhem}}
\def\swedge{{\mem{\wedge}\,}}

\def\da{{\dot{\a}}}
\def\db{{\dot{\b}}}
\def\dc{{\dot{\c}}}

\def\rmA{{\mathrm{A}}}
\def\rmB{{\mathrm{B}}}

\newcommand{\wrap}[1]{{\smash{#1}\vphantom{\beta}}}

\def\lsq{{
    \kern-0.037em
    \adjustbox{scale=0.919,valign=c}{$
        {
            \adjustbox{raise=-0.0855em}{$\lfloor$}
            \llap{\reflectbox{\rotatebox[origin=c]{180}{$\lfloor$}}}
        }
    $}
    \kern-0.04em
}}
\def\rsq{{
    \kern-0.04em
    \adjustbox{scale=0.919,valign=c}{$
        {
            \rlap{\reflectbox{\rotatebox[origin=c]{180}{$\rfloor$}}} 
            \adjustbox{raise=-0.0855em}{$\rfloor$}
        }
    $}
    \kern-0.037em
}}

\def\M{{\mathcal{M}}}

\def\mathe{{\scalebox{1.05}[1.01]{$\mathrm{e}$}}}
\def\sprime{{\mathrlap{\smash{{}^\prime}}{\hspace{0.05em}}}}

\def\i{{\iota}}

\def\tz{{\tilde{z}}}

\newcommand{\cop}[1]{ \mathrlap{c'}\phantom{c}^{\kern0.24em#1} }

\usepackage{tikz}
\usetikzlibrary{calc} 
\usetikzlibrary{shapes.geometric} 
\usetikzlibrary{positioning} 
\usetikzlibrary{fit} 
\usepackage[a]{esvect} 
\tikzset{empty/.style = {inner sep = 0pt, outer sep = 0, minimum size = 0}}
\tikzset{b/.style = {inner sep = 2pt, outer sep = 2pt, minimum size = 12pt}}
\tikzset{s/.style = {inner sep = 2.5pt, outer sep =2.5pt, minimum size = 1pt, font = \small}}
\tikzset{w/.style = {inner sep = 1pt, outer sep = 2pt, minimum size = 12pt, anchor = west}}

\definecolor{sky}{RGB}{144,187,231}
\definecolor{OxyRed}{RGB}{190,70,62}
\definecolor{NitroBlue}{RGB}{91,122,239}
\definecolor{HydrogenLight}{RGB}{245,250,252}

\tikzset{
	line/.style = {draw, line width = 1.1pt, line cap = round, rounded corners = 0.2pt},
	bine/.style = {draw, line width = 1.1pt, line cap = round, rounded corners = 0.0pt, dotted, color=OxyRed},
	dine/.style = {draw, line width = 1.4pt, line cap = round, rounded corners = 0.0pt, double},
	D/.style = {below = -1.1pt, font = \footnotesize\bfseries\sffamily},
	U/.style = {above = -1.2pt, font = \footnotesize\bfseries\sffamily},
	L/.style = {left,  font=\footnotesize},
	R/.style = {right, font=\footnotesize},
	X/.style = {circle, draw=black, fill=HydrogenLight, inner sep=0pt, outer sep=0pt, minimum size=4.5pt, line width=1.1pt},
	Y/.style = {circle, draw=black, fill=NitroBlue, inner sep=0pt, outer sep=0pt, minimum size=4.5pt, line width=1.1pt}
}

\usepackage[export]{adjustbox}

\newcommand{\bb}[1]{\bigg(\,{#1}\,\bigg)}
\newcommand{\BB}[1]{\Big(\,{#1}\,\Big)}
\newcommand{\bigbig}[1]{\big(\mem{#1}\mem\big)}

\newcommand{\bbsq}[1]{\bigg[\,{#1}\,\bigg]}

\newcommand{\lrp}[1]{\left(\mem{#1}\mem\right)}

\def\O{\mathcal{O}}

\def\minie{\tfrac{1}{2}}

\def\bz{\bar{z}}
\def\bZ{\bar{Z}}
\def\bmu{\bar{\mu}}
\def\rambda{\bar{\lambda}}
\def\trambda{\smash{\tilde{\lambda}}}
\def\tmu{\tilde{\mu}}

\newcommand\eq[1]{\begin{align}#1\end{align}}
\newcommand\eqsplit[1]{\begin{align}\begin{split}#1\end{split}\end{align}}
\newcommand\besplit[1]{\begin{split}#1\end{split}}
\newcommand\bealign[2]{\begin{aligned}[#1]#2\end{aligned}}

\def\tz{{\widetilde{z}}}

\def\R{\mathbb{R}}
\def\C{\mathbb{C}}

\def\mflat{\mathbb{M}}
\def\mhat{\smash{\widehat{\mflat}}}
\def\Mhat{\smash{\widehat{\M}}}

\def\tflat{\mathbb{T}}
\def\mt{\mathbb{MT}}
\def\MT{\mathcal{MT}}

\def\kflat{\mathbb{K}}
\def\K{\mathcal{K}}

\def\V{\mathbb{V}}

\def\stheta{\smash{
    \accentset{\adjustbox{scale=0.45}{$\sqrt{}$}}{\theta}\hhnem{}
}}

\def\SO{\mathrm{SO}}
\def\SU{\mathrm{SU}}
\def\GL{\mathrm{GL}}
\def\SL{\mathrm{SL}}

\def\g{\mathfrak{g}}

\def\Cinfty{C^\infty\hnem}

\def\Kerr{{\smash{\text{$\kern-0.075em\sqrt{\text{Kerr\hem}}$}}}}
\def\ga{{\text{G2A}}}

\begin{document}

\title{
	The Kerr Two-Twistor Particle 
}

\author{Joon-Hwi Kim}
\affiliation{Walter Burke Institute for Theoretical Physics, California Institute of Technology, Pasadena, CA 91125}

\begin{abstract}
	An all-orders worldline effective action
	for Kerr black hole
	is achieved in
	twistor particle theory.
\end{abstract}

\preprint{CALT-TH 2026-001}


\bibliographystyle{utphys-modified}

\renewcommand*{\bibfont}{\fontsize{7.5}{7.5}\selectfont}
\setlength{\bibsep}{1pt}

\maketitle

\paragraph{Introduction}%
The twistor particle program
\cite{Penrose:1974di,Perjes:1974ra,%
tod1975dissertation,%
hughston1980programme,
perjes1982introduction,%
perjes1982internal,%
tod1977some,%
tod1976two,%
Perjes:1976sy,penrose1977twistor,
perjes1976evidence,hughston1976twistor,hughston1979twistors,perjes1979unitary%
}
was a movement in the '70s to '80s
that applied twistor theory to particle physics.
In particular, massive particles were implemented as systems of two or more twistors,
whose internal symmetries were associated with color or flavor.

In a modern reboot of this program \cite{ambikerr0,ambikerr1},
the two-twistor implementation is adopted
such that
the $\mathrm{SU}(2)$ internal symmetry
describes the massive little group instead,
in accordance with \rcite{ahh2017}.
This revival has also expanded the scope
from particle physics to
astrophysics,
based on the point-particle effective theory of
macroscopic bodies
\cite{Goldberger:2004jt,Porto:2005ac,Levi:2015msa,Porto:2016pyg,Levi:2018nxp,Kalin:2020mvi}
such as neutron stars or black holes.

The two-twistor system should be able to realize
interacting particles;
see \rrcite{tod1976two,Bette:1989zt,Bette:2004ip,Fedoruk:2007dd,Deguchi:2015iuw,ambikerr1}
for previous attempts.
The problem of direct relevance to the current post-Minkowskian gravity
community
is to implement
the Kerr black hole
coupled to curved spacetime,
capturing all its $2^\ell$-pole moments for $\ell = 0,1,2,\cdots$
\cite{janis1965structure,Newman:1965tw-janis,Newman:1973yu,vines2018scattering,Guevara:2018wpp,Guevara:2019fsj,chkl2019,Levi:2015msa}
as well as further nonlinear-in-curvature couplings.

In this note,
we describe how
an attempt toward curved massive twistor theory
had resulted in
a candidate effective point-particle Lagrangian for the Kerr black hole
to all orders in spin and curvature.
This construction was obtained in 2022
by the present author
and was shared with his colleagues
via private communications,
though its release is unfortunately delayed to this day.

This derivation implements
an intrinsic feature of twistor particle theory
that spin is literally an imaginary deviation
in terms of complexified incidence relation
\cite{Shirafuji:1983zd,penrose:maccallum,newman1974curiosity}.
As a result,
the Newman-Janis algorithm \cite{Newman:1965tw-janis}
is realized
for
dynamical particles.
A differential-geometric formulation
arises 
in terms of
the generator of geodesic deviation, $N$, 
and an almost-complex structure, $J$.

\skip
\paragraph{Free Particle}%
The twistor space $\tflat$ is the K\"ahler vector space $\mathbb{C}^4$
with $(2,2)$-signature metric;
we consider the product $\tflat {\,\times\,} \tflat$
\cite{penrose:maccallum,tod1976two}.
Linear coordinates on $\tflat {\,\times\,} \tflat$ are $Z_\rmA{}^I$.
The index
$\rmA = 0,1,2,3$ is a Dirac spinor
while $I = 0,1$ is an $\mathrm{SU}(2)$ spinor.
The Weyl blocks are
\begin{align}
	Z_\rmA{}^I
	\,=\,
		\Bigg({
		\begin{aligned}
			\lambda_\a{}^I
			\\
			i\hem \mu^{\da I}
		\end{aligned}
		}\Bigg)
	\,,\quad
	\bZ_I{}^\rmA
	\,=\,
	\BB{
		-i\hem \bmu_I{}^\a	
		\,\,\,
		\rambda_{I\da}
	}
	\,,
\end{align}
where $\bZ_I{}^\rmA = [Z_\rmA{}^I]^*$.
The symplectic form is
\eq{
	\label{omega0}
	\omega^\circ \,=\, i\mem d\bZ_I{}^\rmA \swedge dZ_\rmA{}^I
	\,=\,
		d\mu^{\da I}\nem \swedge d\rambda_{I\da}
		+
		d\bmu_I{}^\a\nem \swedge d\lambda_\a{}^I
	\,.
}
The conformal group $\mathrm{SU}(2,2)$ acts from the left
while the massive little group $\mathrm{SU}(2)$ acts from the right.
The $\mathrm{U}(1)$
which $\bZ_I{}^\rmA Z_\rmA{}^I$ generates
is merely gauge
\cite{ambikerr0,ambikerr1}.

Consider the quadric
$-\frac{1}{2}\, I^{\rmA\rmB}\mem Z_\rmA{}^I Z_\rmB{}^J\mem \e_{IJ} = \det(\lambda) = c$
in $\tflat {\,\times\,} \tflat$,
which we denote as $\Delta_c$.
Here, \smash{$I^{\rmA\rmB}$} is the infinity twistor \cite{penrose:maccallum}.
For mathematical precision, one may define
the massive twistor space as
$\mt = (\tflat {\,\times\,} \tflat) {\,\setminus\,} \Delta_0$.
\eqref{omega0} shows that
$\mt$ is
the cotangent bundle
of the space $\lambda_\a{}^I \in \GL(2,\C)$
of \rcite{ahh2017}'s massive spinor-helicity variables.

Symplectic reduction derives
a $12$-dimensional constrained phase space
$\mt_* = \Delta_m / \C$,
realizing the $3$ translational and $3$ rotational degrees of freedom
of a massive spinning body
with rest mass $m>0$
\cite{ambikerr0,ambikerr1}
\footnote{
	It is easy to incorporate the Regge trajectory \cite{Hanson:1974qy};
	see \rrcite{ambikerr0,ambikerr1}.
}.

The Poincar\'e group is included as a subgroup of the $\SU(2,2)$.
The Poincar\'e charges are found as
\eq{
\besplit{
	\label{poincare}
	p_{\a\da}
	\,&=\,
		-\lambda_\a{}^I\hem \rambda_{I\da}
	\,,\\
	j^\da{}_\db
	\,&=\, 
		\mu^{\da I}\hem \rambda_\wrap{I\db}
		- \tfrac{1}{2}\mem \delta^\da{}_\db\,
			\mu^{\dc I}\hem \rambda_\wrap{I\dc}
	\,.
}
}
The Poincar\'e-invariant symplectic potential is
\begin{align}
	\label{theta0Z}
	\theta^\circ \,=\,
	i\,\BB{
		\bZ_I{}^\rmA\mem dZ_\rmA{}^I
		- d\bZ_I{}^\rmA\mem Z_\rmA{}^I
	}
	\,,
\end{align}
such that $\omega^\circ = d\theta^\circ$.

\skip
\paragraph{Twistor Magic}%
A remarkable feature
of the twistor particle framework
is that
an imaginary shift of the two-twistor's co-incident position
derives a spinning particle
from a scalar particle.

As is well-known,
the relation between twistor space and 
spacetime $\mflat = (\R^4,\eta)$ is
established by the incidence relation
\cite{penrose:maccallum,penrose1967twistoralgebra},
the massive adaptation of which reads
\eq{
	\label{incidence}
	\mu^{\da I} 
	\,=\,
		x^{\da\a}\mem \lambda_\a{}^I
	\,.
}
Geometrically, \eqref{incidence} means that the two twistors $Z_\rmA{}^{I=0,1}$
are co-incident at the same spacetime point $x \in \mflat$.
By plugging in \eqref{incidence} to \eqref{theta0Z},
one obtains
\eq{
	\label{free.theta}
	\theta^\circ
	\,=\,
	p_{\a\da}\mem dx^{\da\a}
	\,,
}
which is the symplectic potential
of a scalar particle.

Now consider the complexification of
\eqref{incidence}:
\eq{
	\label{incidencec}
	\mu^{\da I} 
	\,=\,
	\BB{
		x^{\da\a} + i\hem y^{\da\a}
	}\mem \lambda_\a{}^I
	\,.
}
This shifts the co-incident point of the twistors $Z_\rmA{}^{I=0,1}$ to
a point
in complexified Minkowski space
$\mhat = (\C^4,\widehat{\eta})$,
where $\widehat{\eta}$ is the holomorphic extension of the Lorentzian metric $\eta$.
By plugging in \eqref{incidencec} to \eqref{theta0Z},
one obtains
\eq{
\label{free.thetac}
	\theta^\circ
	\,=\,
	p_{\a\da}\mem dx^{\da\a}
	+ i\mem y^{\da\a}\mem
	\BB{
		\lambda_\a{}^I\hem d\rambda_{I\da}
		{\,-\,}
		d\lambda_\a{}^I\mem \rambda_{I\da}
	}
	\,,
}
which is precisely 
the symplectic potential
of a massive spinning particle
\cite{ambikerr0,ambikerr1,tod1976two}.
Inserting \eqref{incidencec}
in \eqref{poincare}
shows that
$y^{\da\a}$ describes
the spin (Pauli-Luba\'nski) pseudovector normalized in the unit of length.
Hence, to put in a well-recognized notation \cite{Hanson:1974qy,Levi:2015msa}
for the reader's sake,
the transition from
\eqref{free.theta} to \eqref{free.thetac} describes
\footnote{
	\label{py}
	Note that the $\mathrm{U}(1)$ gauge generator evaluates to
	$\protect\bZ_I{}^\rmA Z_\rmA{}^I = -p_{\a\protect\da}\mem y^{\protect\da\a}$
	on \protect\eqref{incidencec}.
	In \eqref{rough}, we have imposed $-p_{\a\protect\da}\mem y^{\protect\da\a} = 0$.
}
\eq{
\label{rough}
	p\mem dx
	\mem\,\,\,\,\mapsto\,\,\,\,
	p\mem dx
	+ \tfrac{1}{2}\mem S\mem \Lambda\mem d\Lambda
	\,.
}

\newpage

The insight that the complexified Minkowski space in twistor theory
unifies spacetime $x^\m$ and spin $y^\m$
as real and imaginary parts
traces back to
\rrcite{penrose:maccallum,newman1974curiosity},
whose explicit demonstration for dynamical massless particles
is given in \rcite{Shirafuji:1983zd}.
Here, we have provided the explicit demonstration for dynamical massive particles.

\skip
\paragraph{Spinspacetime}%
The complexified Minkowski space
in this context
is dubbed
spinspacetime \cite{sst-asym}:
\eq{
	\label{tau-flat}
	\mhat
	\,\,\cong\,\,
	T\mflat
	\,.
}
As a real manifold,
spinspacetime is the tangent bundle $T\mflat$ of the spacetime $\mflat$,
with base coordinates $x^{\da\a}$ and fiber coordinates $y^{\da\a}$.
As a complex manifold,
spinspacetime is equipped with complex coordinates
\eq{
	\label{z}
	z^{\da\a} \,=\, x^{\da\a} + iy^{\da\a}
	\,,
}
which are holomorphic
due to the almost-complex structure
$dx^{\da\a} \mapsto dy^{\da\a}$,
$dy^{\da\a} \mapsto -dx^{\da\a}$.
(In this paper, we wish to formulate complex structures as
isomorphisms of the cotangent bundle
rather than of the tangent bundle.)

Smooth tensor fields on spacetime
are analytically continued to
smooth tensor fields on spinspacetime.
Their Cauchy-Riemann flow
arises by the exponentiated Lie derivative
$\exp(\mem i\pounds_N)$,
where $N \in \Gamma(T\mflat)$:
\eq{
	\label{N-flat}
	N \,=\, y^{\da\a}\mem \frac{\partial}{\partial x^{\da\a}}
	\,.
}
In particular,
the complexified metric is
$\widehat{\eta} = \exp(\mem i\pounds_N)\mem \eta$.

As explicated in \rrcite{newman1974curiosity,sst-asym},
the physical origin of \eqref{z} is
a Hodge duality in
the decomposition of angular momentum
for massive particles,
\eq{
	\label{Jsplit}
	j \,=\, (x \mwedge p) + {*}(y \mwedge p)
	\,.
}
The self-dual part of \eqref{Jsplit}
is the self-dual part of $(z \mwedge p)$,
as $*$ is sent to $+i$.
The complexification in \eqref{incidencec}
precisely arises in this way
via $j^\da{}_\db$ in \eqref{poincare}.

\skip
\paragraph{Correspondence Space}%
In massless twistor theory, the incidence relation is mathematically formalized as the double fibration.
Let us achieve its massive equivalent.

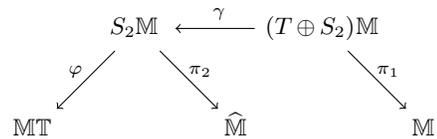
\begin{figure}[t]
	\centering
	\adjustbox{valign=c}{\begin{tikzpicture}
		\node[empty] (o) at (0,0) {};
		\node[empty] (i) at (-1.35, -1.35) {};
		\node[empty] (j) at ( 1.35, -1.35) {};
		\node[empty] (x) at ( 2.5, 0) {};
		\node[b] (A) at ($(o)$) {${
			S_2\mflat
		}\phantom{|}$};
		\node[b] (Ai) at ($(A)+(i)$) {$\mathclap{
			\mt
		}\phantom{|}$};
		\node[b] (Aj) at ($(A)+(j)$) {$\mathclap{
			\mhat
		}\phantom{|}$};
		\node[b] (B) at ($(o)+(x)$) {${
			(T {\,\oplus\,} S_2)\hem\mflat
		}\phantom{|}$};
		\node[b] (Bj) at ($(B)+(j)$) {$\mathclap{
			\mflat
		}\phantom{|}$};
		\draw[->] (A)--(Ai) node[midway,left, pos=0.35] 
		{\scriptsize $\varphi$\,\hnem};
		\draw[->] (A)--(Aj) node[midway,right,pos=0.35] {\scriptsize $\pi_2$};
		\draw[->] (B)--(Bj) node[midway,right,pos=0.35] {\scriptsize $\pi_1$};
		\draw[->] (B)--(A) node[midway,above,pos=0.45] {\scriptsize $\gamma$};
	\end{tikzpicture}}
	\caption{
		Flat massive twistor theory.
	}
	\label{fibration-flat}
\end{figure}

Let $S_2\mhat$ be a trivial $\GL(2,\C)$ bundle over $\mhat$.
\fref{fibration-flat}
concerns two maps
$\varphi: (z^{\da\a};\lambda_\a{}^I) \mapsto (\lambda_\a{}^I , i\hem z^{\da\a}\lambda_\a{}^I)$
and
$\pi_2 : (z^{\da\a};\lambda_\a{}^I) \mapsto z^{\da\a}$
from it.
The first is an invertible diffeomorphism.
The second is the bundle projection.
This provides a mathematical formalization of
the massive incidence relation in \eqref{incidencec},
establishing the relation between
massive twistor space $\mt$ and spinspacetime $\mhat$.

When $\mhat$ is regarded as a real manifold
as per \eqref{tau-flat},
$S_2\mhat$
is viewed as
the direct sum bundle $(T {\,\oplus\,} S_2)\hem\mflat$
over spacetime $\mflat$.
This is represented as a map 
$\gamma: (x^{\da\a};y^{\da\a},\lambda_\a{}^I) \mapsto (x^{\da\a} \mplus iy^{\da\a};\lambda_\a{}^I)$
in \fref{fibration-flat}.

The space
\eq{
	\label{K-flat}
	\kflat
	\mem=\mem
		(T {\,\oplus\,} S_2)\hem\mflat
	\,\,\cong\,\,
		S_2\mhat
}
serves as
the massive analog of the correspondence space,
which has the same dimension as $\mt$.
Crucially, $\kflat$ inherits the K\"ahler geometry of $\mt$
through $\varphi$ and $\gamma$.

First,
$\kflat$ is equipped with an almost-complex structure
\eqsplit{
	\label{J-flat}
	\kern-0.2em
	J
	\,:\,
	(dx^{\da\a},dy^{\da\a},d\lambda_\a{}^I)
	&\mem\mapsto
	(dy^{\da\a},-dx^{\da\a},-i\mem d\lambda_\a{}^I)
	\mem.
	\kern-0.2em
}
This corresponds to the complex structure
$dZ_\rmA{}^I \mapsto -i\mem dZ_\rmA{}^I$ 
of $\mt$.

Second,
$\kflat$ is equipped with the symplectic potential in \eqref{free.thetac}.
This is the pullback of the symplectic potential of $\mt$ given in \eqref{free.theta}
by $\varphi \circ \gamma$.
(By abuse of notation, we have denoted both symplectic potentials as $\theta^\circ$.)

As a result, $\kflat$ is a K\"ahler manifold.
It is left as an exercise to 
derive
the pullback of 
the $(2,2)$-signature metric
by $\varphi \circ \gamma$
and
check the K\"ahler triple relation within $\kflat$.

We also equip $\kflat$
with a vector field $y^{\da\a} \partial/\partial x^{\da\a} \in \Gamma(T\kflat)$,
which is again denoted as $N$ by abuse of notation.

Note that
working in the massive correspondence space $\kflat$
is what \rrcite{Shirafuji:1983zd,Fedoruk:2007dd}
refer to the so-called ``hybrid'' description.
Namely, $\kflat$ is the phase space of a free massive spinning body
described in terms of
spacetime position $x^{\da\a}$, spin length pseudovector $y^{\da\a}$,
and the spin frame $\lambda_\a{}^I$ 
encoding momentum and Eulerian angles.

\skip
\paragraph{\textit{N} and \textit{J} for Newman-Janis}%
The twistor magic can be implemented in the massive correspondence space $\kflat$.
Namely, the spinning-particle symplectic potential
arises from the scalar-particle symplectic potential as
\eq{
\label{magic-in-K}
	\theta^\circ
	\,=\,
	\BB{
		1 + J\mem \i_N\hem d
	}\bigbig{
		p_{\a\da}\mem dx^{\da\a}
	}
	\,.
}
Here, $\i_N$ denotes the interior product
with respect to $N$
while $J$ is the almost-complex structure in \eqref{J-flat}.

\eqref{magic-in-K}
precisely
transcribes
the complexification of the co-incidence relation in \eqref{incidencec}.
Firstly,
$\i_N\hem d$ implements 
a deviation $x \mapsto x + y$
in the manner that conforms to the polarization choice.
Then
$J$ implements $\pm i$ factors
such that the twistors $Z_\rmA{}^I$ are co-incident at $x+iy$
and the dual twistors $\bZ_I{}^\rmA$ are co-incident at $x-iy$.

Note that $J$ implements
Hodge duality on the massive spinor-helicity fibers:
$+i$ for the right-handed $\rambda_{I\da}$
and
$-i$ for the left-handed $\lambda_\a{}^I$.
Physically,
this converts the electric mass dipole
of a deviated scalar particle
to
the magnetic mass dipole,
i.e., spin,
of the spinning particle:
\eq{
	\label{dipoles}
	J\mem\bigbig{
		- dp_{\a\da}\mem y^{\da\a}
	}
	\,=\,
	i\mem y^{\da\a}\mem
		\BB{
			\lambda_\a{}^I\hem d\rambda_{I\da}
			{\,-\,}
			d\lambda_\a{}^I\mem \rambda_{I\da}
		}
	\,.
}
\eqref{dipoles} is an incarnation of the Hodge duality between orbital and spin angular momenta in \eqref{Jsplit}.

\skip
\paragraph{Generalizing to Curved Backgrounds}%
The desire of this paper is to
implement the twistor magic in curved backgrounds.
However,
we do not have
a definition of
``curved massive twistor space'' 
yet.

Recall the scaffolding of curved twistor theory
\cite{penrose1976curvedtwistor,mason2010gravity,mason1989connection}.
One may start by discussing the geometry of the correspondence space
within the conventional mathematical languages of general relativity.
Then the ``radical''
twistor-space picture
will  arise afterwards
via the self-duality condition
as
integrability along $\a$-surfaces.

We shall follow the same approach.
First of all,
we have to realize
the curved massive correspondence space $\K$
as
the typical phase space of a massive spinning particle
in textbook general relativity.
After establishing this ``conservative'' description,
we may then envision how 
a notion of
curved massive twistor space $\MT$ could emerge.

\skip
\paragraph{Curved Correspondence Space}%
In textbook general relativity,
spacetime $(\M,g)$ is a real-analytic pseudo-Riemannian four-manifold.
Local trivialization of $T\M$
describes the 
frame $E_m = E^\m{}_m(x)\mem \partial_\m$
and
coframe $e^m = e^m{}_\m(x)\mem dx^\m$
such that
$\langle e^m , E_n \rangle = \delta^m{}_n$.
Here, $m,n,\cdots = 0,1,2,3$ are local Lorentz indices
while $\m,\n,\cdots = 0,1,2,3$ are spacetime indices.

The decomposition
$T\M \cong (S^+ \mtensor S^-)\M$
into self-dual and anti-self-dual spinor bundles
describes that
a local Lorentz index is a pair of spinor indices
as
$e^m \leftrightarrow e^{\da\a}$.
The Levi-Civita connection $\nabla$ of $g$
splits to $\SL(2,\C)$-valued connections $\bgamma$ and $\gamma$ on
$S^\pm\hnem\M$.
The structure equations are
\smash{$
	0 = 
		De^{\da\a}
	=
		de^{\da\a} - \bgamma^\da{}_\db \swedge e^{\db\a} + e^{\da\b} \swedge \gamma_\b{}^\a
$}
and
\smash{$
	R_\a{}^\b =
		d\gamma_\a{}^\b + 
		\gamma_\a{}^\c \wedge \gamma_\c{}^\b
$},
where
$D$ denotes the covariant exterior derivative
with respect to $\nabla$.

Physically,
the degrees of freedom of
a massive spinning particle in $(\M,g)$
are described by
$x^\m$, $y^m$, and $\lambda_\a{}^I$.
Here, the index form declares the behaviors
under coordinate and local Lorentz transformations.
The position variable $x^\m$ is supposed to describe coordinates on $\M$.
The spin length pseudovector $y^m$ 
and 
the spin frame $\lambda_\a{}^I$
are supposed to be local Lorentz degrees of freedom.

Therefore, the curved massive correspondence space
should be defined as the direct sum bundle
\eq{
	\label{K-curved}
	\K 
	\,=\,
		(T {\,\oplus\,} S_2)\hem\M
	\,,
}
equipped
with base coordinates $x^\m {\,\in\,} \R^4$
and fiber coordinates 
$y^m {\,\in\,} \R^4$,
$\lambda_\a{}^I {\,\in\,} \GL(2,\C)$.
Here,
$S_2\M$ is a principal $\GL(2,\C)$-bundle over $\M$
which is isomorphic to an open subbundle of $(S^- \moplus S^-)\M$.

The transition functions for $\K$ 
as a manifold
are restricted to the following specific forms:
\eqsplit{
\label{transfs}
	(x^\m,y^m,\lambda_\a{}^I)
	&\,\,\mapsto\,\mem
	(f^\m(x),y^m,\lambda_\a{}^I)
	\,,\\
	(x^\m,y^m,\lambda_\a{}^I)
	&\,\,\mapsto\,\mem
	(x^\m,
		\Omega^m{}_n(x)\, y^n,
		\Omega_\a{}^\b\hnem(x)\, \lambda_\a{}^I
	)
	\,.
}
These encode
spacetime diffeomorphisms
and local Lorentz transformations,
respectively.
Here,
$\Omega^m{}_n(x) \in \SO(1,3)$
arises from
$\Omega_\a{}^\b(x) \in \SL(2,\C)$.

The curved massive correspondence space
is equipped with 
a triple of geometrical structures:
\eq{
	(\K,N,J,\omega)
	\,.
}
$N {\,\in\,} \Gamma(T\K)$ is a vector field,
$J : T^*\K {\,\to\,} T^*\K$ is an almost-complex structure,
and $\omega {\,\in\,} Z^2(\K) {\:\subset\:} \Omega^2(\K)$ is a symplectic structure.
$N$ and $J$
are invariant under the restricted coordinate transformations of $\K$
stipulated in \eqref{transfs}.

First,
$N$ is
uniquely defined by the interior products
\eq{
	\label{Ndef}
	\i_N
	\,\,\,:\,\,\,
	(e^m,Dy^m,D\lambda_\a{}^I)
	\,\,\mapsto\,\,
	(y^m,0,0)
	\,.
}
It is easy that
$N = y^m \tilde{E}_m$,
where $\tilde{E}_m \in \Gamma(T\K)$
is
the horizontal lift \cite{ehresmann1948connexions,Mason:2013sva} of
the frame vector field
$E_m \in \Gamma(T\M)$
with respect to $\nabla$.
It follows that $N$ is the generator of geodesic deviation
and transport
\cite{gde}.

Second, $J$ is uniquely defined as the map such that
\eqsplit{
	\label{J}
	J
	\,\,:\,\,
	(e^m,Dy^m,D\lambda_\a{}^I)
	&\,\hhem\mapsto\mem
	(Dy^m,-e^m,-i\mem D\lambda_\a{}^I)
	\mem.
}

Third,
$\omega \in Z^2(\K)$ is defined as a symplectic form
that approaches to the following two-form $\omega^\bullet \in \Omega^2(\K)$
in the limit of vanishing curvature of $\nabla$,
namely $\lim_{R\to0} \omega = \omega^\bullet$:
\eqsplit{
\label{cov.omega}
	\omega^\bullet
	\,=\,
	{}&{}
	\rambda_{I\da}\mem \bigbig{
		e^{\da\a} \mminus i\mem Dy^{\da\a}
	} \wedge D\lambda_\a{}^I
	\\
	{}&{}
	- D\rambda_{I\da} \wedge \bigbig{
		e^{\da\a} \mplus i\mem Dy^{\da\a}
	}\mem \lambda_\a{}^I
	\\ 
	{}&{}
	+ 2i\mem y^{\da\a}\mem
		D\lambda_\a{}^I \swedge D\rambda_{I\da}
	\,.
}
Here, we define
$p_{\a\da} {\:=\:} {-\lambda_\a{}^I\hem \rambda_{I\da}}$.

It should be clear that
$N$ in \eqref{Ndef},
$J$ in \eqref{J},
and $\omega^\bullet$ in \eqref{cov.omega}
respectively
reduce to
$N$ in \eqref{N-flat},
$J$ in \eqref{J-flat},
and
the exterior derivative of \eqref{free.thetac}
in the limit of vanishing curvature of $\nabla$,
in which case $\K \cong \kflat$.
They are the very covariantizations of
the free theory's
$N$, $J$, and $\omega^\circ = d\theta^\circ$
due to $\nabla$.

When $\nabla$ is curved,
$\omega$ must differ from $\omega^\bullet$
since $d\omega^\bullet \neq 0$.
In this sense the particle's symplectic structure
necessarily develops a curvature correction.

The structures $N$ and $J$ 
could also develop curvature corrections in principle,
but
it will suffice to deform just the symplectic structure
for our physical purposes.

The particle's $12$-dimensional physical phase space
arises by
symplectic reduction of $\K$ 
by
gauge generators
$-p_m\hhem y^m \approx 0$ (see Footnote\:\cite{Note1})
and $p^2 \mplus m^2 \approx 0$.

\skip
\paragraph{Traditional Minimal Coupling}%
By choosing the symplectic structure $\omega$ differently,
one realizes various gravitational couplings of the massive spinning particle
in the phase space $\K$.
In particular, the choice of $\omega$ 
amounts to specifying the particle's action.

For example, the choice
dubbed ``minimal coupling''
in the traditional sense
defines the symplectic potential on $\K$
by directly covariantizing \eqref{free.thetac}:
\eq{
\label{cov.theta}
	\theta_{(0)}
	\,=\,
	p_m\hhem e^m
	+ i\mem y^{\da\a}\mem
	\BB{
		\lambda_\a{}^I\hem D\rambda_{I\da}
		{\,-\,}
		D\lambda_\a{}^I\mem \rambda_{I\da}
	}
	\,.
}
In this case, 
the symplectic form is $\omega_{(0)} = d\theta_{(0)} = \omega^\bullet + \smash{\omega'_{(0)}}$,
where 
\smash{$\omega'_{(0)} = p_m\hem {\star\hnem}R^m{}_n\hem y^n$}.
$\star$ denotes the Hodge star acting on the internal (local Lorentz) indices.
Note that it could be instructive to rewrite \eqref{cov.theta} as
\eq{
	\theta_{(0)}
	\,=\,
		p_m\hhem e^m
		+ p_m {\hem\star\Theta}^m{}_n\hem y^n + W_0\mem d\psi
	\,,
}
where $\Theta^m{}_n$ is a Lorentz-valued Maurer-Cartan form
such that $D\Theta^m{}_n {\,=\,} R^m{}_n$,
$\psi$ is a $\mathrm{U}(1)$ angle,
and $W_0 {\,=\,} {-p_m\hhem y^m}$.

\begin{figure*}[t]
	\centering
	\begin{tikzpicture}
	    \node[empty] (O) at (0,0) {};
	    \node[empty] (X) at (4.4, 0) {};
	    \node[empty] (x) at (3.9, 0) {};
	    \node[empty] (Y) at (0, -0.92) {};
	    \node[w] (a00) at ($(O)$) {$p_me^m$};
	    \node[w] (a01) at ($(O)+1*(X)$) {${d(p_my^m)}$};
	    \node[w] (a02) at ($(O)+2.0*(X)$) {$0$};
	    \node[w] (a10) at ($(O)+1*(Y)$) {$-Dp_m\hem y^m$};
	    \node[w] (a11) at ($(O)+1*(Y)+1*(X)$) {$0$};
	    \node[w] (a20) at ($(O)+2*(Y)$) {$p_m (\i_NR^m{}_n)\hem y^n$};
	    \node[w] (a21) at ($(O)+2*(Y)+1*(X)$) {$0$};
	    \node[w] (a30) at ($(O)+3*(Y)$) {$p_m (\i_ND\mem \i_NR^m{}_n)\hem y^n$};
	    \node[w] (a31) at ($(O)+3*(Y)+1*(X)$) {$0$};
	    \node[w] (a40) at ($(O)+4*(Y)$) {$\vdots$};
	    \node[w] (a2K) at ($(O)+2*(Y)-1*(X)$) {$p_m {\hem\star\Theta}^m{}_n\hem y^n {\mem+\mem} W_0\mem d\psi$};
	    \node[w] (a3K) at ($(O)+3*(Y)-1*(X)$) {$p_m (\i_N{\star R}^m{}_n)\hem y^n$};
	    \node[w] (a4K) at ($(O)+4*(Y)-1*(X)$) {$p_m (\i_ND\mem \i_N{\star R}^m{}_n)\hem y^n$};
	    \node[w] (a5K) at ($(O)+5*(Y)-1*(X)$) {$\vdots$};
	    \node[w] (a2k) at ($(O)+2*(Y)-1*(X)+(x)$) {$0$};
	    \node[w] (a3k) at ($(O)+3*(Y)-1*(X)+(x)$) {$0$};
	    \node[w] (a4k) at ($(O)+4*(Y)-1*(X)+(x)$) {$0$};
	    \node[w] (phantom-a00) at ($(O)$) {};
	    \node[w] (phantom-a01) at ($(O)+1*(X)$) {};
	    \node[w] (phantom-a02) at ($(O)+1.5*(X)$) {};
	    \node[w] (phantom-a10) at ($(O)+1*(Y)$) {};
	    \node[w] (phantom-a11) at ($(O)+1*(Y)+1*(X)$) {};
	    \node[w] (phantom-a20) at ($(O)+2*(Y)$) {};
	    \node[w] (phantom-a21) at ($(O)+2*(Y)+1*(X)$) {};
	    \node[w] (phantom-a30) at ($(O)+3*(Y)$) {};
	    \node[w] (phantom-a31) at ($(O)+3*(Y)+1*(X)$) {};
	    \node[w] (phantom-a40) at ($(O)+4*(Y)$) {};
	    \node[w] (phantom-a2K) at ($(O)+2*(Y)-1*(X)$) {};
	    \node[w] (phantom-a3K) at ($(O)+3*(Y)-1*(X)$) {};
	    \node[w] (phantom-a4K) at ($(O)+4*(Y)-1*(X)$) {};
	    \node[w] (phantom-a5K) at ($(O)+5*(Y)-1*(X)$) {};
	    \node[w] (phantom-a2k) at ($(O)+2*(Y)-1*(X)+(x)$) {};
	    \node[w] (phantom-a3k) at ($(O)+3*(Y)-1*(X)+(x)$) {};
	    \node[w] (phantom-a4k) at ($(O)+4*(Y)-1*(X)+(x)$) {};
	    \draw[->] (a00)--(a01) node[midway,above] {\scriptsize $d\mem\i_N$};
	    \draw[->] (a01)--(a02) node[] {};
	    \draw[->] (a10)--(a11) node[] {};
	    \draw[->] (a20)--(a21) node[] {};
	    \draw[->] (a30)--(a31) node[] {};
	    \draw[->] (phantom-a00)--(phantom-a10) node[midway,left] {\scriptsize $\i_N d$};
	    \draw[->] (phantom-a10)--(phantom-a20) node[] {};
	    \draw[->] (phantom-a20)--(phantom-a30) node[] {};
	    \draw[->] (phantom-a01)--(phantom-a11) node[] {};
	    \draw[->] (phantom-a10)--(a2K) node[midway,above] {\scriptsize $J$};
	    \draw[->] (phantom-a2K)--(phantom-a3K) node[] {};
	    \draw[->] (phantom-a3K)--(phantom-a4K) node[] {};
	    \draw[->] (a2K)--(a2k) node[] {};
	    \draw[->] (a3K)--(a3k) node[] {};
	    \draw[->] (a4K)--(a4k) node[] {};
	    \draw[->] (phantom-a30)--(phantom-a40) node[] {};
	    \draw[->] (phantom-a4K)--(phantom-a5K) node[] {};
	\end{tikzpicture}
	\caption{
		The ``$\i_N d / J$ sequence'' for Kerr.
		A tree of one-forms 
		emanates from
		the scalar-particle symplectic potential, $p_m\hhem e^m$.
	}
	\label{tree-of-life}
\end{figure*}

The resulting Hamiltonian equations of motion are
immediate by
the technique of covariant symplectic perturbations.
They reproduce the the Mathisson-Papapetrou-Dixon \cite{Mathisson:1937zz,Papapetrou:1951pa,Dixon:1970zza}
equations,
which describe the quadrupolar coupling
(gravimagnetic ratio \cite{Yee:1993ya,khriplovich1989particle,Khriplovich:1997ni})
$C_2 {\,=\,} 0$.
However,
it is well-known that
the Kerr black hole carries $C_2 {\,=\,} 1$
\cite{janis1965structure,Newman:1965tw-janis,Newman:1973yu,vines2018scattering,Guevara:2018wpp,Guevara:2019fsj,chkl2019}.
Therefore,
the symplectic potential in \eqref{cov.theta}
fails to describe the Kerr black hole.

\skip
\paragraph{\textit{N} and \textit{J} for Newman-Janis, with Curvature}%
In spirit of the Newman-Janis algorithm,
we ask whether
the Kerr symplectic potential
can be constructed in $\K$
solely from
the Schwarzschild symplectic potential $p_m\hem e^m$
and
the invariant structures $N$ and $J$.

\eqref{magic-in-K} has implemented the twistor magic
in the flat correspondence space $\kflat$.
Its covariantization is
\eq{
\label{theta0}
	\theta_{(0)}
	\,=\,
	\BB{
		1 + J\mem \i_N\hem d
	}\bigbig{
		p_m\hhem e^m
	}
	\,,
}
which unfortunately evaluates to \eqref{cov.theta}.
A natural modification is
\eq{
\label{theta1}
	\theta_{(1)}
	\,=\,
	\BB{
		\cos(\pounds_N)
		+
		\sinc(\pounds_N)\,
		J\mem \i_N\hem d
	}\bigbig{
		p_m\hhem e^m
	}
	\,.
}

\eqref{theta1}
faithfully implements the concept of Newman-Janis algorithm:
imaginary shifts along spin direction.
Since $N$ is the generator of geodesic deviation,
the imaginary deviations arise by
$\exp(\pm i\pounds_N)
= \cos(\pounds_N) \pm i \sinc(\pounds_N)\hem \pounds_N
$.
\eqref{theta1} implements
the $\pm i$ factors with the almost-complex structure $J$
while replacing a $\pounds_N$ with $\i_N\hem d$
as a direct generalization of
the flat construction in \eqref{magic-in-K}.

Crucially, the repeated action of $\pounds_N$
does not truncate at a finite order
in the presence of curvature.
As per the Cartan magic formula,
the repeated action of $\pounds_N = \i_N\hem d + d\hem \i_N$
is diagrammatically represented as
a planar grid
where each node is a differential form
and each edge is an arrow representing
the action of either
$\i_N\hem d$ or $d\hem \i_N$,
such that $(\i_N\hem d)(d\hem \i_N) = 0$
and $(d\hem \i_N)(\i_N\hem d) = 0$.
As shown in \fref{tree-of-life},
this grid does not truncate
if $R^m{}_n \neq 0$.

Hence we shall regard that $\cos(\pounds_N)$ and $\sinc(\pounds_N)$
were secretly present in \eqref{magic-in-K} as well,
albeit trivialized.
In curved backgrounds,
they ensure the unity $C_\ell {\:=\:} 1$ of even and odd multipoles coefficients.
That is,
\eqref{theta1}
correctly captures the linear gravitational coupling of Kerr
established through
\rrcite{janis1965structure,Newman:1965tw-janis,Newman:1973yu,vines2018scattering,Guevara:2018wpp,Guevara:2019fsj,chkl2019}.

In fact, \eqref{theta1}
pinpoints a unique gravitational coupling
to all orders in spin and curvature.
From \fref{tree-of-life},
it can be seen that
\eqref{theta1} can be represented as
\eq{
	\label{sum1}
	\theta_{(1)}
	\mem&=\,
	\theta_{(0)}
	+\mem\hhem
		\sum_{\ell=2}^\infty
			\frac{1}{\ell!}\,
				p_m
				\bigbig{\hnem
					(\i_N\hem D)^{\ell-2} \i_N\mem {\star^\ell\hnem} R^m{}_n
				\hhnem}\hem y^n
	\,,
}
where ${\star^\ell}$ means to act on the internal Hodge star $\ell$ times.
By methods in differential geometry \cite{gde},
\eqref{sum1} is explicitly computed as
\vspace{-1.0\baselineskip}
\begin{widetext}
\vspace{-1.0\baselineskip}
\vspace{-0.35\baselineskip}
\eq{
\label{theta-earth}
	\theta_{(1)}
    	\mem=\,
    		\theta_{(0)}
    	 	+\mem\hhem \sum_{\ell=2}^\infty
    	 		\frac{1}{\ell!}
    			\sum_{p=1}^{\lfloor{\ell/2}\rfloor}\kern-0.2em
    			\sum_{\a \in \Omega_p(\ell)}\kern-0.2em
    	 			\bbsq{
    	 				\prod_{i=1}^p
    	 				\binom{
    	 					\bigbig{
    	 						\sum_{j=i}^p \a_j
    	 					} \mminus 2
    	 					\hem
    	 				}{
    	 					\a_i \mminus 2
    	 				}
    	 				\nem\nem
    	 			}\mem
    	 		\lrp{
    	 		\bealign{c}{
    	 			&
				\bigbig{\hnem
					({\star}^\ell Q_{\a_1}\hnem)\mem Q_{\a_2}\, {\nem\cdots\mem} Q_{\a_p}
				\hnem}
				{}^m{}_s\, e^s
				\\
				&
				+ (\a_p\mminus2)\mem
				\bigbig{\hnem
					({\star}^\ell Q_{\a_1}\hnem)\mem Q_{\a_2}\, {\nem\cdots\mem} Q_{\a_{p-1}} Q_{\a_p - 1}
				\hnem}
				{}^m{}_s\, Dy^s
			\hnem
    	 		}}
  	 \,,
}
\vspace{-0.55\baselineskip}
\end{widetext}
\phantom{.}

\vspace{-2.1\baselineskip}\noindent
where $\a = (\a_1,\a_2,\cdots,\a_p) \in \Omega_p(\ell)$
runs over ordered partitions
such that
$\a_1 + \a_2 + \cdots + \a_p  = \ell$
and
$\a_i \geq 2$.
The so-called $Q$-tensors \cite{gde} are defined as
\begin{align}
    \label{Qtensor}
    ({\star}^\ell Q_j)^m{}_s
    \,=\,
	  {\star}^\ell R^m{}_{r_1r_2s;r_3;\cdots;r_j}\hnem(x)
	  \, y^{r_1}{\cdots}y^{r_j}
    \,.
\end{align}

The nonlinear-in-Riemann terms
in \eqref{theta-earth}
are fine-tuned to yield
remarkable physical consequences
in self-dual backgrounds,
as we have explained in \rcite{probe-nj}.

\phantom{.}

\vspace{-2.0\baselineskip}
\paragraph{Heavenly Portal to Curved Spinspacetime}%
\eqref{theta1} and its explicit evaluation in \eqref{theta-earth}
provide a manifestly covariant, gauge-invariant, and reparametrization-invariant
construction of the Kerr action
in generic, real spacetimes.
It is based on two invariant geometric structures in the spinning-particle phase space:
$N$ and $J$.

However, it is based on the typical languages of general relativity.
Apparently,
the spinspacetime and massive twistor descriptions are lost.
This means that the curved analogs of the left part of the diagram in \fref{fibration-flat}
are yet to be constructed.

The first step toward the ``radical'' reformulations
is to identify the map $\gamma$.
It can be discovered by
traveling to self-dual spacetimes,
i.e., \textit{heaven}
in the terminologies of
Newman \cite{shaviv1975general,Newman:1976gc} and Pleba\'nski \cite{plebanski1975some,Plebanski:1977zz}.

The spacetime curvature $R^m{}_n$
splits to self-dual and anti-self-dual parts as
\smash{$\bR^\da{}_\db$} and \smash{$R_\a{}^\b$}.
Consider the (formal) complexified limit in which the latter vanishes,
so $\star R^m{}_n = +i\hem R^m{}_n$.
Then \eqref{sum1} equals
\eq{
	\label{kerr.heaven}
	\theta_{(1)}
	\mem&=\,
	\theta_{(0)}
	+\mem\hhem
		\sum_{\ell=2}^\infty
			\frac{i^\ell}{\ell!}\,
				p_m
				\bigbig{\hnem
					(\i_N\hem D)^{\ell-2} \i_N R^m{}_n
				\hhnem}\hem y^n
	\,.
}
Notably, \eqref{kerr.heaven} is nearly equal to
the symplectic potential of a scalar particle,
geodesically deviated in a pure-imaginary direction $iy$
(cf. \rcite{gde}):
\eq{
	\label{sch+ia}
	&
	\mathe^{i\pounds_N}\bigbig{
		p_m\hhem e^m
	}
	\\[-0.2\baselineskip]
	&
	{={}}\,
		p_m\hhem e^m
		+ i\mem p_m Dy^m
		+\mem\hhem
			\sum_{\ell=2}^\infty
				\frac{i^\ell}{\ell!}\,
					p_m
					\bigbig{\hnem
						(\i_N\hem D)^{\ell-2} \i_N R^m{}_n
					\hhnem}\hem y^n
	\,.
	\nonumber
}
That is, \eqrefs{kerr.heaven}{sch+ia}
differ only by the combination
\eq{
	\label{g2a}
	\theta_\ga
	\,=\,
		p_m\mem {\star}\Theta^m{}_n\hem y^n
		- i\mem p_m Dy^m
		+ W_0\mem d\psi
	\,,
}
precisely since
the self-dual limit facilitates
replacing $J$ in \eqref{theta1}
with $+i$
except at the dipolar spin order $\ell = 1$.

Therefore,
our Kerr particle
is indistinguishable from
an imaginary-deviated Schwarzschild particle
in self-dual backgrounds,
up to a slight subtlety at the dipolar order.

\begin{figure}[t]
	\centering
	\includegraphics[scale=0.88]{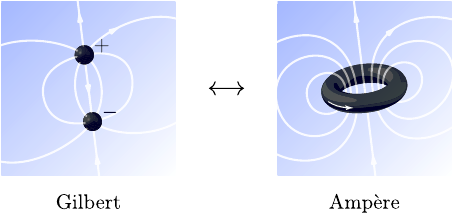}
	\caption{
		Gilbert-Amp\`ere duality.
	}
	\label{fig:G2A}
\end{figure}

To understand 
the origin of
this subtlety
from a physicist's intuition,
note that \eqref{g2a} replaces
an imaginary \textit{electric mass dipole} term $i\mem p_m Dy^m$
with 
the \textit{magnetic mass dipole} (i.e., spin) term $p_m\mem {\star}\Theta^m{}_n\hem y^n$.
This reflects a kinematics-level difference between
spinning and non-spinning particles.
Specifically,
there is a physical sense in which
\eqref{g2a}
replaces a Gilbertian dipole
(displaced charges)
with an Amp\`erian dipole
(current loop),
as illustrated in \fref{fig:G2A}.
In self-dual backgrounds,
magnetic charge is indistinguishable from $i$ times electric charge.
Thus the Amp\`erian dipole
$p_m\mem {\star}\Theta^m{}_n\hem y^n$
gets replaced with $i$ times the Gilbertian dipole $p_m Dy^m$,
and upon this very mechanism
spin becomes $i$ times deviation---%
which is the gist of the Newman-Janis algorithm, in fact \cite{nja}.

Hence,
physically speaking, 
the almost-complex structure $J$
has implemented a duality between Gilbertian and Amp\`erian multipoles.
This observation can be rephrased as
electric-magnetic or Hodge duality;
recall the discussions around \eqrefs{magic-in-K}{dipoles}.

With this insight,
we rewrite the Kerr symplectic potential
in the self-dual limit as
\eqsplit{
	\label{kerr.g2a}
	\theta_{(1)}
	\mem&=\,
	\mathe^{i\pounds_N}\bigbig{
		p_m\hhem e^m
	}
	\hem+\mem\hhem \theta_\ga
	\,,\\
	\mem&=\,
	\mathe^{i\pounds_N}\bigbig{
		p_m\hhem e^m
		\hem+\mem\hhem \theta_\ga
	}
	\,,
}
where the second line follows from the fact that
$\theta_\ga$ in \eqref{g2a}
carries no gravitational charge
in the self-dual limit
such that
$d\theta_\ga = 0$;
it describes a sort of a topological defect.

Remarkably, \eqref{kerr.g2a} 
shows that
the Kerr symplectic potential
in self-dual spacetime
is 
a one-form
localized at a ``complex spacetime point''
whose coordinates are
\eq{
	\label{zcurved}
	\mathe^{iN} x^\m
	\,=\,
		x^\m 
		+ i\mem y^\m
		+ \tfrac{1}{2}\, \Gamma^\m{}_{\r\s}(x)\mem y^\r y^\s
		+ \O(y^3)
	\,.
}
\eqref{zcurved} generalizes \eqref{z}
in a nonlinear and generally covariant fashion:
\textit{curved spinspacetime}.

\begin{figure}[t]
	\centering
	\adjustbox{valign=c}{\begin{tikzpicture}
		\node[empty] (o) at (0,0) {};
		\node[empty] (i) at (-1.35, -1.35) {};
		\node[empty] (j) at ( 1.35, -1.35) {};
		\node[empty] (x) at ( 3.0, 0) {};
		\node[b] (A) at ($(o)$) {${
			\K
		}\phantom{|}$};
		\node[b] (Ai) at ($(A)+(i)$) {$\mathclap{
			\MT
		}\phantom{|}$};
		\node[b] (Aj) at ($(A)+(j)$) {$\mathclap{
			\M_z
		}\phantom{|}$};
		\node[b] (B) at ($(o)+(x)$) {${
			(T^{1,0} {\,\oplus\,} S_2^{1,0} {\,\oplus\,} S_2^{1,0})\hem\M
		}\phantom{|}$};
		\node[b] (Bj) at ($(B)+(j)$) {$\mathclap{
			\M_x
		}\phantom{|}$};
		\draw[->] (A)--(Ai) node[midway,left, pos=0.35] 
		{\scriptsize $\varphi$\,};
		\draw[->] (A)--(Aj) node[midway,right,pos=0.35] {\scriptsize $\pi_2$};
		\draw[->] (B)--(Bj) node[midway,right,pos=0.35] {\scriptsize $\pi_1$};
		\draw[->] (B)--(A) node[midway,above,pos=0.44] {\scriptsize $\gamma$};
	\end{tikzpicture}}
	\caption{
		Curved massive twistor theory.
	}
	\label{fibration-curved}
\end{figure}
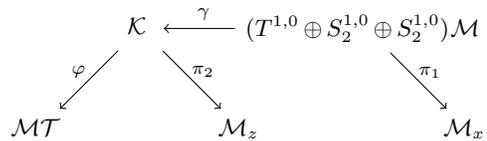

\skip
\paragraph{Curved Spinspacetime}%
As shown in Appendix A,
spinspacetime exists for any
real-analytic spacetime.
To expedite our discussion, however,
we may want to directly jump to a complexified setup
regarding its relevance to the self-dual limit and twistorial ideas.

From now on, spacetime 
is a complex-analytic four-manifold $(\M,g)$
equipped with a holomorphic metric $g$.

Spinspacetime is the holomorphic tangent bundle $\Mhat = T^{1,0}\M$,
which has base coordinates $x^\m$ and fiber coordinates $y^\m$.
This means that coordinate transformations on $\M$
induce
$(x^\m,y^\m) \mapsto (f^\m(x),f^\m{}_{,\n}(x)\mem y^\n)$,
where $f^\m(x)$ are holomorphic functions.

Spinspacetime is equipped with the holomorphic vector field
$N \in T^{1,0}\Mhat$
that is horizontal with respect to the Levi-Civita connection $\nabla$ of $g$
and satisfies $\i_N dx^\m = y^\m$.

The time-$i$ flow of $N$ is
a map $\gamma : \Mhat \to \Mhat$:
\eqsplit{
	z^{\m\sprime}
	\,=\,
		\delta^{\m\sprime}{}_\m\mem
		(\mathe^{iN} x^\m)
	\,,\quad
	y^{\m\sprime}
	\,=\,
		\delta^{\m\sprime}{}_\m\mem
		(\mathe^{iN} y^\m)
	\,.
}
Here, we are adopting the primed index notation
of Synge calculus
\cite{Ruse:1931ht,Synge:1931zz,Synge:1960ueh,Poisson:2011nh,gde}.
$z^{\m\sprime}$ are the coordinates of the point in $\M$
that is reached by geodesically flowing from $x$ along the direction of $y$,
so
$\Exp : T^{1,0}\M \to \M : (x,y) \mapsto z$
is the holomorphic exponential map.
$y^{\m\sprime} {\:=\:} W^{\m\sprime}{}_\m\mem y^\m$ describe the tangent to the geodesic segment
at the endpoint $z$,
where $W^{\m\sprime}{}_\m$ is the parallel propagator from $x$ to $z$.

In sum, we identify two coordinate charts on $\Mhat = T^{1,0}\M$.
The ``defining'' chart is $(x^\m;y^\m)$;
the other chart \smash{$(z^{\m\sprime};y^{\m'})$}
will be referred to as a ``googly'' chart.

\newpage
\paragraph{Curved Correspondence Space}%
For the complex spacetime $(\M,g)$,
the curved massive correspondence space
is the holomorphic direct sum bundle
\eq{
	\label{K-curved-holic}
	\K 
	\,=\,
		(T^{1,0} {\,\oplus\,} S_2^{1,0} {\,\oplus\,} S_2^{1,0})\hem\M
	\,,
}
equipped
with base coordinates $x^\m {\,\in\,} \C^4$
and fiber coordinates 
$y^m {\,\in\,} \C^4$,
$\lambda_\a{}^I, \trambda_{I\da} {\,\in\,} \GL(2,\C)$.
Its transition functions are restricted 
in terms of holomorphic functions
$f^\m(x) {\:\in\:} \C^4$
and
$\Omega_\a{}^\b(x), \tilde{\Omega}^\da{}_\db(x) {\:\in\:} \SL(2,\C)$
(cf. \eqref{transfs}).

The curved correspondence space is equipped with
the holomorphic vector field
$N {\:\in\:} \Gamma(T^{1,0}\K)$
that is horizontal with respect to $\nabla$
and satisfies $\i_N dx^\m {\:=\:} y^\m$,
which we again denote as $N$ by abuse of notation.

Similarly,
the time-$i$ flow of $N$ is denoted as $\gamma$.
This is
the map $\gamma : \K \to \K$
such that
$x^\m \mapsto z^{\m\sprime} = (\Exp(x,y))^{\m\sprime}$,
$y^\m \mapsto y^{\m\sprime} {\:=\:} W^{\m\sprime}{}_\m\mem y^\m$,
and
$\lambda_\a{}^I \mapsto \lambda_{\a'}{}^I {\:=\:} W_{\a'}{}^\a\mem \lambda_\a{}^I$,
$\trambda_{I\da} \mapsto \trambda_{I\da\sprime} {\:=\:} \trambda_{I\da}\mem W^\da{}_{\da'}$.
Here,
the parallel propagator \smash{$W^{\m\sprime}{}_\m$}
decomposes into $\SL(2,\C)$ blocks.
Note that it
is metric-preserving as
$g_{\m\sprime\n'}(z)\mem W^{\m\sprime}{}_\m\mem W^{\n\sprime}{}_\n
{\:=\:} g_{\m\n}(x)$.

The defining chart on $\K$ is
$(x^\m;y^\m,\lambda_\a{}^I,\trambda_{I\da})$.
The googly chart on $\K$ is
$(z^{\m\sprime};y^{\m\sprime},\lambda_{\a'}{}^I,\trambda_{I\da'}\hnem)$.

The former chart establishes the projection
$\pi_1 : \K \to \M : (x^\m;y^\m,\lambda_\a{}^I,\trambda_{I\da}) \mapsto x^\m$.
The latter chart establishes the projection
$\pi_2 : \K {\:\to\:} \M : (z^{\m\sprime};y^{\m\sprime},\lambda_{\a'}{}^I,\trambda_{I\da'}\hnem) {\:\mapsto\:} z^{\m\sprime}$.
As a result, $\K$ fibers on $\M$ in two different ways.
If a disambiguation is needed, we denote the former and latter versions of $\M$ as 
$\M_x$ and $\M_z$, respectively.

Note that $\Mhat = T^{1,0}\M_x \cong \M_z \mtimes \M_\tz$,
where $z$ and $\tz$ arise by $\mathe^{+iN}x$ and $\mathe^{-iN}x$.
In the flat limit $\M_x {\,\to\:} \mflat$,
$\M_z$ and $\M_\tz$
are precisely the holomorphic and anti-holomorphic
subspaces of $\mhat = \mflat^\C$,
$\mhat^{1,0}$ and $\mhat^{0,1}$.

Lastly,
$\K$ is not only equipped with the horizontal vector field
$N {\,\in\,} \Gamma(T^{1,0}\K)$
but also
\smash{$J : T^*_{1,0}\K {\,\to\,} T^*_{1,0}\K$}
such that $J^2 {\:=\:} {-\mathrm{id}}$
and $\omega {\,\in\,} \Omega^{2,0}(\K)$
such that $d\omega {\:=\:} 0$:
$(\K,N,J,\omega)$.
$N$ and $J$
are invariant under the restricted coordinate transformations of $\K$
stipulated above.

The Kerr symplectic potential 
arises by the same formula in \eqref{theta1},
yet now $\theta^{(1)} \in \Omega^{1,0}(\K)$.

\skip
\paragraph{Chiral Action and Motion in Heaven}%
If the holomorphic metric $g$ is self-dual,
then \eqref{kerr.g2a} implies that
the Kerr symplectic potential is
$\theta_{(1)} = \gamma^* \theta_{(1)}^+ + d( iW_0 )$,
where
\eq{
	\label{chiral.theta}
	\theta_{(1)}^+
	\,=\,
		- \trambda_{I\da'}\hem e^{\da\sprime\a}\hem \lambda_\a{}^I
		- 2i\mem \trambda_{I\da'}\mem y^{\da\sprime\a}\mem
			d\lambda_\a{}^I
	\,.
}
For simplicity, we have set the anti-self-dual spin connection coefficients to zero in a local patch
so that the anti-self-dual Wilson line $W_{\a'}{}^\b$ is trivial;
hence the primes on left-handed spinor indices have been dropped.

\eqref{chiral.theta} provides a chiral formulation of the Kerr particle,
treating $\lambda_\a{}^I$ and $\trambda_{I\da'}$ asymmetrically.
This formulation makes it evident that the symplectic form
is isomorphic to the free theory's symplectic form $\omega^\circ$
via the map
$\varrho: 
(dz^{\da\a}, d\bz^{\da\a}, d\lambda_\a{}^I , d\rambda_{I\da})
\mapsto
(e^{\da\sprime\a} , e^{\da\sprime\a} - 2iDy^{\da\sprime\a}, d\lambda_\a{}^I, D\trambda_{I\da'})
$.
Hence
there is no covariant symplectic perturbation,
and the resulting Hamiltonian equations of motion
are just isomorphic to free theory:
\eqsplit{
	\label{heavenly-eom}
	e^{m\sprime}{}_{\m'}(z)\mem \dot{z}^{\m\sprime} \mem=\mem p^{m\sprime}
	&\,,\quad
	\dot{\lambda}_\a{}^I \mem=\mem 0
	\,,\\
	Dy^{m\sprime}\nem/d\t
		\mem=\mem 0
	&\,,\quad
	D\trambda_{I\da\sprime}\mem/d\t
		\mem=\mem 0
	\,.
}
Remarkably,
the dynamics of the Kerr black hole in heaven
is thus simply geodesic motion
with parallel transportation of the local Lorentz degrees of freedom
$y^{m\sprime}, \trambda_{I\da'}, \lambda_\a{}^I$
on the complex worldline $z^{\m\sprime}$
\footnote{
	A similar idea seems to had been conceived in \rcite{ko1981theory}
	in the context of $\mathscr{H}$-space.
}.

\eqref{heavenly-eom} concretely verifies the spinning, complexified analog of equivalence principle
that uniquely characterizes spinning black holes in self-dual external fields,
which the current author proposed through works \cite{sst-asym,ambikerr1}:
the left-handed spin frame $\lambda_\a{}^I$ is frozen, i.e., does not precess at all,
in self-dual backgrounds.

When transcribed to the defining chart
(covariantized at $x^\m$ in spacetime),
\eqref{heavenly-eom}
yields
an all-orders-in-spin extension of 
Mathisson-Papapetrou-Dixon equations
with $C_\ell {\:=\:} 1$ for all $\ell$.
Moreover,
\eqref{heavenly-eom}
implies exact hidden symmetries
in the presence of Killing and Killing-Yano tensors
by dynamically Newman-Janis shifting the conserved quantities of the Schwarzschild probe
\cite{probe-nj}.

It is also easily seen that the holomorphic coordinates $z^{\m\sprime}$ of 
the Kerr black hole's
curved spinspacetime $\Mhat$ remain Poisson-commutative
in self-dual gravity.

\skip
\paragraph{Curved Massive Twistor Space}%
Lastly, we expand the holomorphic coframe 
around a flat background as
$e^{\da\sprime\a} = dz^{\da\sprime\a} + h^{\da\sprime\a}$.
The physical relevance of this setup is perturbation theory, for instance.
Concretely, employ the second Pleba\'nski coordinates
\cite{plebanski1975some,Adamo:2021bej}
to let $\a$ remain as the ``rigid'' $\SL(2,\C)$ index.
Then \eqref{chiral.theta} becomes
\eq{
	\label{chiral.theta.h}
	\kern-0.5em
	{- \trambda_{I\da'}\hem d\bigbig{
		z^{\da\sprime\a}\hem \lambda_\a{}^I
	}}
	+ \trambda_{I\da'}
	\bigbig{
		z^{\da\sprime\a} {\hem-\mem} 2iy^{\da\sprime\a}
	}\hem d\lambda_\a{}^I
	+ \theta'_{(1)}
	\,,
	\kern-0.6em
}
where 
$\theta' = - \trambda_{I\da'}\hem h^{\da\sprime\a}\hem \lambda_\a{}^I
	= p_{m'} h^{m\sprime}
	= p_{m'} h^{m\sprime}{}_{\m'}(z)\mem dz^{\m\sprime}
$.
This suggests to identify the ``mu variables'' as
\eqsplit{
	\label{muvariables}
	\mu^{\da\sprime\mem I}
	\,=\,
		z^{\da\sprime\a}\hem\hhem \lambda_\a{}^I
	\,,\quad
	\tmu_I{}^\a
	\,&=\,
		\trambda_{I\da'}
		\bigbig{
			z^{\da\sprime\a} {\hem-\mem} 2iy^{\da\sprime\a}
		}
	\,,
}
in which case \eqref{chiral.theta.h} 
and its exterior derivative become
\eqsplit{
	\label{twistor.h}
	\theta^+_{(1)}
	\,&=\,
	- \trambda_{I\da'}\hem d\mu^{\da\sprime\mem I}
	+ \tmu_I{}^\a\hem d\lambda_\a{}^I
	+ \theta'_{(1)}
	\,,\\
	\omega_{(1)}
	\,&=\,
	d\mu^{\da\sprime\mem I} \swedge d\trambda_{I\da'}
	+ d\tmu_I{}^\a \swedge d\lambda_\a{}^I
	+ d\theta'_{(1)}
	\,.
}

The free-theory part of \eqref{twistor.h}
is identical to $\omega^\circ$ in \eqref{omega0}
upon unpriming the right-handed spinor indices.
\eqref{twistor.h}---eventually---verifies the Kerr symplectic structure in self-dual spacetime we derived in \rcite{ambikerr1}
from a bootstrap approach based on the K\"ahler geometry (zig-zag structure) of massive twistor space
and the Souriau insight
\cite{souriau1970structure,dyson1990feynman}
that
deformations of particle symplectic structure
implement couplings to external fields.

\eqref{twistor.h}
establishes that 
the coupling of the Kerr particle to self-dual gravity
can be formulated as
holomorphic deformations of the symplectic structure on massive twistor space:
\eq{
	(\mt,\omega^\circ)
	\,\,\,\xrightarrow{\,\,\,\,\,\,}\,\,\,
	(\MT,\omega_{(1)})
	\,.
}
By holomorphic deformations, we mean that the holomorphic subspace of $\mt$ remains as a Lagrange submanifold
after altering the symplectic structure to $\omega_{(1)}$.

In this context,
\eqref{muvariables} is identified
as the deformed massive incidence relation,
from which the scaffolding of curved massive twistor theory arises.
Namely, \eqref{muvariables} defines a diffeomorphism
$\varphi: \K \to \MT$
and completes the map of our journey shown in \fref{fibration-curved}.

We may note that
$\varphi$ seems to encode a deformation of complex structure as well
when approached from the holomorphic complex structure of curved spinspacetime $\Mhat$.
The original complex coordinates
of $\Mhat \cong \M_z {\,\times\,} \M_\tz$
are $z^{\m\sprime}$ and \smash{$\tz^{\m'\nem\sprime}$}
that describe $\mathe^{\pm iN}x$.
However, the deformed massive incidence relation in \eqref{muvariables}
identifies
$z^{\m\sprime} \mminus 2iy^{\m\sprime}$
as the ``$z$-bar'' variable:
\eq{
	\label{tz}
	w^{\m\sprime}
	\,=\,
		z^{\m\sprime} - 2iy^{\m\sprime}	
	\,=\,
		z^{\m\sprime} - 2i\mem \s^{\m\sprime}(z,\tz)
	\,.
}
Here, $\s(z,\tz)$ describes the complexified Synge world function 
\cite{Ruse:1931ht,Synge:1931zz,Synge:1960ueh,Poisson:2011nh},
$\s : \M_z \mtimes \M_\tz \to \C$.
Crucially,
$w^{\m\sprime}$ differs from \smash{$\tz^{\m'\nem\sprime}$}
for nonzero curvature.

In massless curved twistor theory,
the celebrated nonlinear graviton construction of Penrose \cite{penrose1976curvedtwistor}
formulates self-dual gravity
as deformations of the complex structure in twistor space.
This describes deformations of the twistor lines
in terms of the deformed incidence relation.

Our proposal above
envisions curved massive twistor theory
from a deformed massive incidence relation
relevant to perturbation theory,
which is reminiscent yet not fully equivalent to massless curved twistor theory.
As sketched in Appendix B,
we have also considered a formulation
that follows the ideas of massless curved twistor theory
more closely,
in which case 
the symplectic structure might be fixed.

\skip
\paragraph{Googly Kerr Action}%
In \eqref{theta-earth},
we have already obtained the Kerr action
in generic spacetimes with both self-dual and anti-self-dual modes,
i.e., \textit{earth}
in the terminologies of
Newman \cite{shaviv1975general,Newman:1976gc} and Pleba\'nski \cite{plebanski1975some,Plebanski:1977zz}.

We have then took an excursion to self-dual backgrounds, i.e., \textit{heaven},
to discover the nonlinear Newman-Janis shift in \eqref{zcurved}.
The key punchline is that
the \textit{Newman-Janis shift manifests in heaven}
as the localization of the Kerr action on the holomorphic worldline $z^{\m\sprime}$.

It should be stressed that
there is an obstruction in simultaneously manifesting
the Newman-Janis shifts in earth for both self-dual and anti-self-dual sectors.
This obstruction is due to the very nonlinearity of gravity
and poses a massive analog of the googly 
\cite{Penrose:2015lla:Palatial,penr04-googly,Witten:2003nn,Adamo:2017qyl} problem.
Concretely, it is impossible to write the earthly Kerr action
as a sum of two separate worldline actions localized respectively at
$z^{\m\sprime}$ and $\tz^{\m'\nem\sprime}$,
since there exist nonlinear mixing between self-dual and anti-self-dual curvatures.

See also Appendix C,
where the same problem is found for the case of nonabelian gauge theory.

The best we can do is to perturb away from the self-dual sector,
which is a typical twistor-theorist move.
Assuming the complexified setup,
the Kerr symplectic potential in 
\eqref{theta1}
can be written as
\eq{
\label{wsform}
	\theta_{(1)}
	\,=\,
		p_m\mem e^m
	+
		\frac{
			\mathe^{i\pounds_N} \mminus 1
		}{\pounds_N}
		\,\theta^+_\ga
	+
		\frac{
			\mathe^{-i\pounds_N} \mminus 1
		}{\pounds_N}
		\,\theta^-_\ga
	\,,
}
where we have defined
\eq{
\label{strings.kerr}
	\theta^\pm_\ga
	\mem:=\mem
		\frac{1}{2}\,
		\BB{
			p_m Dy^m
			\mp i\mem
			\bigbig{
				p_m\mem {\star}\Theta^m{}_n\hem y^n
				\mplus W_0\mem d\psi
			}
		\nem}
	\,.
}
Meanwhile, the symplectic potential
$\mathe^{i\pounds_N}\bigbig{p_m\hhem e^m}$
of the imaginary-deviated Schwarzschild particle is
\eq{
\label{strings.sch+ia}
	p_m\hhem e^m
	+
		\frac{
			\mathe^{i\pounds_N} \mminus 1
		}{\pounds_N}
		\,\theta^+_\ga
	+
		\frac{
			\mathe^{i\pounds_N} \mminus 1
		}{\pounds_N}
		\,\theta^-_\ga
	\,.
}
The difference between \eqrefs{strings.kerr}{strings.sch+ia} derives
\vspace{-1.0\baselineskip}
\begin{widetext}
\vspace{-1.0\baselineskip}
\vspace{-0.35\baselineskip}
\eq{
\label{kerr.googly}
	\theta_{(1)}
	\mem&=\,
	\mathe^{i\pounds_N}\mem
	\bb{
		p_m\hhem e^m
		+ 
		\frac{
			\mathe^{-2i\pounds_N} \mminus 1
		}{\pounds_N}
		\,\theta^-_\ga
	}
	\,=\,
		\mathe^{i\pounds_N}\mem
		\bb{
			\theta_{(0)}
			- i\mem p_m Dy^m
			+\mem\hhem
				\sum_{\ell=2}^\infty
					\frac{(-2i)^\ell}{\ell!}\mem
						p_m
						\bigbig{\hnem
							(\i_N\hem D)^{\ell-2} \i_N\mem R^-{}^m{}_n
						\hhnem}\hem y^n
		}
	\nonumber
	\,,\\
    	\mem&=\,
\begin{aligned}[t]
		&
		p_{m'}\hhem e^{m\sprime}
		+ p_{m'}\mem {\star}\Theta^{m\sprime}{}_{n'}\hem y^{n\sprime}
		- i\mem p_{m'} Dy^{m\sprime}
		+ W_0\mem d\psi
    		\\
    		&
    	 	+\mem\hhem \sum_{\ell=2}^\infty
    	 		\frac{(-2i)^\ell}{\ell!}
    			\sum_{p=1}^{\lfloor{\ell/2}\rfloor}\kern-0.2em
    			\sum_{\a \in \Omega_p(\ell)}\kern-0.2em
    	 			\bbsq{
    	 				\prod_{i=1}^p
    	 				\binom{
    	 					\bigbig{
    	 						\sum_{j=i}^p \a_j
    	 					} \mminus 2
    	 					\hem
    	 				}{
    	 					\a_i \mminus 2
    	 				}
    	 				\nem\nem
    	 			}\mem
    	 		\lrp{
    	 		\bealign{c}{
    	 			&
				\bigbig{\hnem
					(Q^-_{\a_1}\hnem)\mem Q_{\a_2}\, {\nem\cdots\mem} Q_{\a_p}
				\hnem}
				{}^{m\sprime}{}_{s'}\mem e^{s\sprime}
				\\
				&
				+ (\a_p\mminus2)\mem
				\bigbig{\hnem
					(Q^-_{\a_1}\hnem)\mem Q_{\a_2}\, {\nem\cdots\mem} Q_{\a_{p-1}} Q_{\a_p - 1}
				\hnem}
				{}^{m\sprime}{}_{s'}\mem Dy^{s\sprime}
    	 		}}
 	 \,,
\end{aligned}
}
\vspace{-0.55\baselineskip}
\end{widetext}
\phantom{.}

\vspace{-2.1\baselineskip}\noindent
where 
$R^\pm := \minie\mem (R {\,\mp\,} i\mem {\star} R)$
and
$(Q_j^\pm) := \minie\mem (Q_j {\,\mp\,} i\mem {\star} Q_j)$.
To be precise, the first line is the pullback of the second line by the diffeomorphism $\gamma : \K \to \K$.

\eqref{kerr.googly} provides
a googly formulation of the Kerr action.
It describes
\textit{both} self-dual and anti-self-dual couplings
as an action localized on the \textit{holomorphic} worldline.
It reduces to \eqref{kerr.g2a} in the self-dual limit.
It treats self-dual and anti-self-dual modes,
and the $z^{\m\sprime}$ and \smash{$\tz^{\m'\nem\sprime}$} worldlines,
asymmetrically.

In actual applications,
\eqref{kerr.googly} facilitates
a perturbation theory
in which the Newman-Janis shift (as spin exponentiation  \cite{ahh2017,Guevara:2018wpp,Guevara:2019fsj,chkl2019})
of scattering amplitudes
is manifestly guaranteed in the self-dual sector
to all graviton multiplicities
\cite{probe-nj},
on top of which
the anti-self-dual perturbations are systematically added on.

\skip
\paragraph{Interpretation}%
In \eqref{kerr.googly},
$(\mathe^{-2i\pounds_N} \mminus 1)/(-2i\pounds_N) = 
\smash{\int_0^1\mem d\eta\, \mathe^{-2i\eta\pounds_N}}
$
inserts the Newman-Janis shifts of $\theta^-_\ga$
along the geodesic segment
joining the two points $z$ and $\tz$.
Hence \eqref{kerr.googly} states that
the Kerr black hole 
describes 
a dangling open string attached to
a point mass at $z$.
This string is charged under anti-self-dual gravitons
($d\theta^-_\ga {\:=\:} p_m R^-{}^m{}_n\hem y^n$)
but is invisible to self-dual fields
as remarked earlier.
In fact,
recalling our earlier discussions on Gilbertian and Amp\`erian dipoles,
it precisely describes the anti-self-dual combination
of electric and magnetic mass dipole moments.
Thus, we identify its interpretation as
an anti-self-dual Misner string.

Similarly,
it can be seen that
\eqref{theta1}
describes two point masses
joined by a thin gravitomagnetic flux tube:
an ordinary Misner string
\footnote{
	This string structure was noticed in
	\rcite{gmoov},
	but the physical interpretations and googly formulation
	are not provided there.
	\rcite{gmoov} also does not provide explicit expressions 
	such as Eqs.\:(\ref{theta-earth}) and (\ref{kerr.googly})
	that expand out the action fully in terms of covariant worldline operators.
}.
The two pictures,
googly and real,
are equivalent
via identifying a pair of opposite-sign masses
as a line defect of electric mass dipole.

Misner strings are
the gravitational analogs of the Dirac string;
readers unfamiliar with the concept
may consult to
\rrcite{Misner:1963flatter,Bonnor:1969ala,sackfield1971physical,Mazur:1986gb,Griffiths_Podolský_2009_Ch3,note-sdtn}.

The above analysis elicits a provocative idea that
the Kerr black hole describes a system of self-dual and anti-self-dual dyons.
At least in the effective theory of Kerr black holes,
\eqrefs{kerr.g2a}{kerr.googly} show that
this statement holds true as the physical origin of the Newman-Janis shift.
The question is whether it holds in the ultraviolet theory,
i.e., general relativity,
as a nonperturbative fact.

The persistent endeavor of the current author around this question
had resulted in works \cite{note-sdtn,nja},
which established that the Kerr metric
represents the exact nonlinear superposition of
self-dual and anti-self-dual Taub-NUT solutions
as a holomorphic saddle,
from which the Newman-Janis algorithm \cite{Newman:1965tw-janis}
is faithfully reproduced.

The worldlines
$z^{\m\sprime}$ and \smash{$\tz^{\m'\nem\sprime}$} 
are the infrared avatars of the Taub-NUT instantons.

\begin{figure}[t]
	\centering
	\adjustbox{valign=c}{
		\includegraphics[scale=0.9
		,clip=true,trim=7pt 7pt 7pt 7pt,
		rotate=180
		]{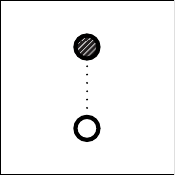}
	}
	$\xleftrightarrow[\textsc{\scriptsize{morphic}}]{\,\,\textsc{\scriptsize{Diffeo-}}\,\,}$\,\,
	\mem\,
	\adjustbox{valign=c}{
		\includegraphics[scale=0.9
		,clip=true,trim=7pt 7pt 7pt 7pt
		]{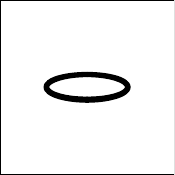}    
	}
	\caption{
		The ring singularity of Kerr black hole
		turns into
		a pair of
		self-dual and anti-self-dual
		Taub-NUT instantons.
	}
	\label{fig:factorization}
\end{figure}

The worldline actions due to
\eqrefs{theta1}{kerr.googly}
insert the Misner string along the geodesic
between the two instantons.
Currently,
there is
no established evidence
that the Misner string
is a topological surface defect
in dynamical, full (non-self-dual) gravity,
although the current author had developed some ideas
while initiating the work \cite{GenSymGrav}.
Hence it is not clear
to the current author
whether the Misner worldsheet can take any shape
to predict the same physical observables.

If the geodesicity of the Misner string
can be assumed,
it seems that \eqref{theta1},
equivalently \eqref{kerr.googly},
defines the unique answer
for the effective action 
that describes
a pair of self-dual and anti-self-dual Taub-NUT instantons.

\skip
\paragraph{Conclusions}%
In this note,
we derived an explicit all-orders worldline effective action for Kerr black hole
by generalizing an intrinsic feature of twistor particle theory
in curved spacetime.
This implements
the Newman-Janis algorithm at the level of test-particle actions,
in terms of differential-geometric elements $N$ and $J$.
Nonlinear Newman-Janis shift manifests in self-dual backgrounds
as localizations on the holomorphic worldline.
The full gravitational interactions
are systematically understood by
perturbing around the self-dual sector.
Clear physical interpretations are given
and are substantiated by 
the ultraviolet description
of the Kerr black hole.

\newpage

This note is written in the geometrical language
and wishes to deliver itself to the twistor theory community as well.
This note attempts to make
a small yet ambitious step towards
``curved massive twistor theory,''
which shall exist
in light of the histories of massless twistor theory 
and twistor particle program.
Some tentative identifications and definitions are given for
the deformed massive incidence relation
and curved massive twistor space,
based on the particular case of the Kerr black hole
as a massive twistor.

\section*{Appendix}

\paragraph{A. Spinspacetime from Adapted Complex Structure}%
This appendix elaborates on the definition of spinspacetime
for real spacetime.

In \eqref{tau-flat},
we discussed
two perspectives toward
spinspacetime:
real and complex manifolds.
In the case of curved real-analytic spacetime $(\M,g)$,
however,
the latter picture
is seemingly lost.
The formula in \eqref{z}
becomes utterly nonsensical,
since
$x^\m$ are \textit{coordinates}
while $y^\m$ are components of a \textit{vector}.
This represents a  clash between
two necessary features
of ``curved spinspacetime'':
general covariance and holomorphy.

Luckily,
the mathematical framework
known as adap\-ted complex structure
\cite{guillemin1991grauert,guillemin1992grauert,lempert1991global,szHoke1991complex,halverscheid2002complexifications,aguilar2001symplectic,burns2000symplectic,hall2011adapted}
provides a nice resolution,
showing that any tangent bundle $T\M$
of a real-analytic $(\M,g)$
can be promoted to a complex manifold $\Mhat$
via complexifying the exponential map
(geodesic flow):
\eq{
	\label{tau}
	\Pi_i
	\,\,:\,\,
	T'\hnem\M
	\,\,\to\,\,
	\Mhat
	\,.
}
Here, we have supposed 
a sufficiently narrow neighborhood $T'\hnem\M$ of the zero section in $T\M$.

The idea is simple.
The time-$\eta$ exponential map
sends the pair $(x,y) {\:\in\:} T'\hnem\M$ to 
the point in $\M$
whose coordinates are
$w^\m(x,y;\eta)
= x^\m
	+ \eta\mem y^\m
	- \smash{\frac{\eta^2}{2}}\, \Gamma^\m{}_{\r\s}(x)\mem y^\r y^\s
	+ \O(y^3)
$.
This is a
power series solution
which converges in $T'\hnem\M$.
Upon a real-analytic coordinate transformation $x^\m \mapsto f^\m(x)$ on $\M$,
the coordinates
of the geodesically deviated point
transform as 
$w^\m(x,y;\eta)
\mapsto f^\m(w(x,y;\eta))
$
for any $\eta$,
by construction.

By analytic continuation,
$z^\m(x,y) {\,:=\,} w^\m(x,y;i)$
is well-defined within $T'\hnem\M$,
whose explicit formula
is exactly \eqref{zcurved}.
By analyticity,
the coordinate transformation on $\M$
induces
$z^\m(x,y)
\mapsto f^\C{}^\m(z(x,y))
$,
where $f^\C$ is the analytic continuation of $f$.
This shows that
$z^\m(x,y)$
defines a holomorphic coordinate chart on $T'\hnem\M$.
Therefore, $T'\hnem\M$ becomes a complex manifold.

Physically speaking,
this construction ensures that
holomorphy is an observer-independent notion.
This is important since holomorphy will be linked with self-duality
by the Newman-Janis shift.

Now let us unravel the precise mathematical details.
The time-$\eta$ exponential map can be formalized as
\eq{
	\Pi_\eta
	\,=\,
		\pi \circ \Phi_\eta
	\,\,:\,\,
	T'\hnem\M \,\to\, \M
	\,,
}
where $\Phi_\eta : T'\hnem\M \to T\M$ denotes the time-$\eta$ flow
by the horizontal vector field $N \in \Gamma(T\M)$
such that $\i_N dx^\m = y^\m$,
and $\pi : T\M \to \M$ is the bundle projection.
The pullback ${\Pi_\eta}^*$
maps
analytic functions on $\M$
to
analytic functions on $T'\hnem\M$,
the explicit power series solutions for which
arise by
the exponentiated Lie derivative
$\mathe^{\eta\pounds_N}$.

By analytic continuation to $\eta \in \C$,
one defines ${\Pi_i}^*$ as a map from
real-analytic functions on $\M$
to complex-analytic functions on $T'\hnem\M$,
the application of which on coordinate functions
establishes 
the map $\Pi_i$ in \eqref{tau}
such that $\M$, as the zero section of $T\M$,
is embedded in the complex-analytic manifold $\Mhat = \M^\C$
as a totally real submanifold of maximal dimension
\cite{whitney1959quelques,grauert1958levi,hall2011adapted}.

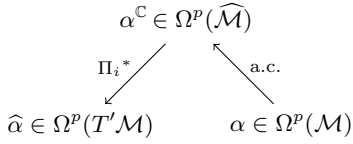
\begin{figure}[t]
	\centering
	\adjustbox{valign=c}{\begin{tikzpicture}
		\node[empty] (o) at (0,0) {};
		\node[empty] (i) at (-1.4, -1.4) {};
		\node[empty] (j) at ( 1.4, -1.4) {};
		\node[b] (A) at ($(o)$) {${
			\a^\C \nem\in \Omega^p(\Mhat)
		}\phantom{|}$};
		\node[b] (Ai) at ($(A)+(i)$) {$\mathclap{
			\widehat{\a} \in \Omega^p(T'\hnem\M)
		}\phantom{|}$};
		\node[b] (Aj) at ($(A)+(j)$) {$\mathclap{
			\a \in \Omega^p(\M)
		}\phantom{|}$};
		\draw[->] (A)--(Ai) node[midway,left,pos=0.4] 
		{\scriptsize ${\Pi_i}^*$\nem};
		\draw[<-] (A)--(Aj) node[midway,right, pos=0.4] 
		{\scriptsize \,\text{a.c.}};
	\end{tikzpicture}}
	\caption{
		The ``central dogma'' of 
		adapted complex structure.
	}
	\label{dogma}
\end{figure}

The practical use of this formalism
is facilitated by the fact that
the exponentiated Lie derivative $\mathe^{i\pounds_N}$
computes the pullback ${\Pi_i}^*$.
Any differential $p$-form $\a {\:\in\:} \Omega^p(\M)$
is transcribed to
$\a^\C \nem{\:\in\:} \smash{\Omega^p(\Mhat)}$
by analytic continuation,
which translates to
$\widehat{\a} = {\Pi_i}^*[\mem{ \a^\C }\mem] = \mathe^{i\pounds_N} \a \in \Omega^p(T'\hnem\M)$.
This is the ``central dogma'' of adapted complex structure,
depicted in \fref{dogma}.
For instance,
a zero-form
$\phi \in \Cinfty(\M)$
exhibits
the covariant Taylor expansion
\eqsplit{
	\label{taylor0}
	\widehat{\phi}(x,y)
	\,&=\,
		\phi^\C(z(x,y))
	\,=\,
		\mathe^{i\pounds_N} \phi(x)
	\,,\\
	\,&=\,
		\sum_{\ell=0}^\infty\,
			\frac{i^\ell}{\ell!}\,
				\phi_{;\r_1;\cdots;\r_\ell}(x)\,
			y^{\r_1} {\cdots} y^{\r_\ell}
	\,.
}
Note how general covariances coexist 
at both points $x$ and $z$
\eqref{taylor0}.

Curved spinspacetime 
is the complex-analytic four-manifold $\Mhat$
that recasts the tangent bundle of spacetime
via adapted complex structure,
in the sense of \eqref{tau}.

Spacetime fields,
such as $\phi(x)$ in \eqref{taylor0},
permeate into spinspacetime
via analytic continuation:
$\phi^\C(z)$.
The spinspacetime field $\phi^\C(z)$
is represented in the ``spacetime $+$ spin'' form,
\smash{$\widehat{\phi}(x,y)$},
by means of the operator
$\mathe^{i\pounds_N}$
\footnote{
	The latter description is perturbative in spin
	and may break down beyond 
	the region
	$T'\hnem\M$
	in $T\M$.
	However, 
	it suffices to consider small enough spin
	for the physical purposes of this paper.
}.

Historically, 
it was Newman who first envisioned
the concept of spinspacetime
and its physical applications
\cite{%
	newman1988remarkable,newman1974curiosity,newman1974collection,%
	Newman:1973yu,Newman:1973afx,Newman:2002mk,Newman:2004ba,%
	Newman:1976gc,ko1981theory,grg207flaherty%
}.
The curved spinspacetime constructions of
this paper transparently realizes Newman's provisional insights
in terms of concrete differential-geometric constructions.

\skip
\paragraph{B. Deformed Incidence Relation}%
This appendix attempts an alternative definition of
curved massive twistor space.

In massless twistor theory,
the deformed incidence relation
takes the form $\mu^\da {\:=\:} F^\da(x,\lambda)$,
where \smash{$d_x F^\da(x,\lambda) = \Lambda^\da{}_\db(x,\lambda)\mem e^{\db\b}\hem \lambda_\b$}
for some $\SL(2,\C)$ frame \smash{$\Lambda^\da{}_\db(x,\lambda)$}
\cite{mason2010gravity}.
Here, $d_x$ is the exterior derivative fixing $\lambda_\a$ in the correspondence space.
Concretely,
$F^\da(x,\lambda)$
is constructed
in Pleba\'nski's second heavenly coordinates
as $F^\da(x,\lambda) = x^{\da\a}\mem \lambda_\a + \Phi^\da(x)\mem ({\lambda_0}^2\nem/\lambda_1) + \cdots$
\cite{plebanski1975some,dunajski2000hyper,Adamo:2021bej}.

Given this construction,
\eqref{chiral.theta} can be written as
\eqsplit{
	\theta_{(1)}^+
	\,=\,
	- \trambda'_{I\da'}\mem d\mu'^{\da\sprime\mem I}
	+ \tmu'_I{}^\a\mem d\lambda_\a{}^I
	\,,
}
where
\begin{subequations}
\eq{
	\smash{\trambda'_{0\da'}(z,\lambda^0,\trambda_0)}
	\,&=\,
		\smash{\trambda_\wrap{0\db'}\mem \L^{\db\sprime}{}_\wrap{\da'}(z,\lambda^0)}
	\,,\\
	\smash{\trambda'_{1\da\sprime\,}(z,\lambda^1,\trambda_1)}
	\,&=\, 
		\smash{\trambda_\wrap{1\db'}\mem \L^{\db\sprime}{}_\wrap{\da'}(z,\lambda^1)}
	\,,\\
	\smash{\mu'^{\da\sprime\, 0}(z,\lambda^0)}
	\,&=\,
		\smash{F^{\da\sprime}(z,\lambda^0)}
	\,,\\
	\smash{\mu'^{\da\sprime\, 1}(z,\lambda^1)}
	\,&=\,
		\smash{F^{\da\sprime}(z,\lambda^1)}
	\,,\\
	\tmu_I{}^\a(z,\tz,\lambda,\trambda)
	\,&=\,
		\trambda'_\wrap{J\db'}\,
			\frac{\partial \mu'^{\db\sprime\mem J}}{\partial \lambda_\a{}^I}
		-2i\mem \trambda_\wrap{I\db'}\hem
			\s^{\db\sprime\a}(z,\tz)
	\,.
}
\end{subequations}

This approach fixes the symplectic structure of $\mt$
while altering the definitions of
$\mu^{\da I}$, $\rambda_{I\da}$, and $\bmu_I{}^\a$.
However, 
its physical relevance seems rather unclear.
It seems to break
covariance under the $\SU(2)$ massive little group,
due to the nonlinear dependence of $F^\da(x,\lambda)$ on $(\lambda_0/\lambda_1)$.

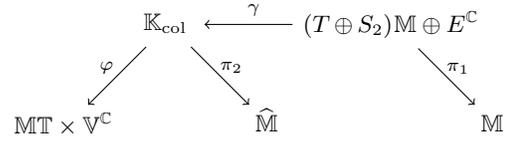
\begin{figure}[t]
	\centering
	\adjustbox{valign=c}{\begin{tikzpicture}
		\node[empty] (o) at (0,0) {};
		\node[empty] (i) at (-1.35, -1.35) {};
		\node[empty] (j) at ( 1.35, -1.35) {};
		\node[empty] (x) at ( 3.0, 0) {};
		\node[b] (A) at ($(o)$) {${
			\kflat_\text{col}
		}\phantom{|}$};
		\node[b] (Ai) at ($(A)+(i)$) {$\mathclap{
			\mt \times \V^\C
		}\phantom{|}$};
		\node[b] (Aj) at ($(A)+(j)$) {$\mathclap{
			\mhat
		}\phantom{|}$};
		\node[b] (B) at ($(o)+(x)$) {${
			(T {\,\oplus\,} S_2)\hem\mflat
			\oplus E^\C
		}\phantom{|}$};
		\node[b] (Bj) at ($(B)+(j)$) {$\mathclap{
			\mflat
		}\phantom{|}$};
		\draw[->] (A)--(Ai) node[midway,left, pos=0.35] 
		{\scriptsize $\varphi$\,};
		\draw[->] (A)--(Aj) node[midway,right,pos=0.35] {\scriptsize $\pi_2$};
		\draw[->] (B)--(Bj) node[midway,right,pos=0.35] {\scriptsize $\pi_1$};
		\draw[->] (B)--(A) node[midway,above,pos=0.44] {\scriptsize $\gamma$};
	\end{tikzpicture}}
	\caption{
		Massive twistor theory with a color extension.
	}
	\label{fibration-gauge}
\end{figure}

\skip
\paragraph{C. {\Kerr} Action}%
This appendix constructs the {\Kerr} symplectic structure coupled to nonabelian gauge theory.
Suppose gauge group $G$
and gauge algebra $\g$.
Let $a,b,\cdots$ be the adjoint indices.
Let $\V$ be a representation space
assigned with indices $i,j,\cdots$,
on which the generators are $(t_a)^i{}_j$.

The color degrees of freedom
can be implemented by phase spaces such as
$T^*G$, $T^*\V$, etc.
Take $T^*\V$ with complex coordinates
$\psi^i$, for instance.
The correspondence space is enlarged as $\kflat_\text{col} = (T{\:\oplus\:}S_2)\mflat \oplus E$,
where $E$ is a vector bundle over $\mflat$
whose typical fiber is $T^*\V$.

Let $A {\:\in\:} \Omega^1(\mflat;\g)$ be the nonabelian gauge connection.
A complete, gauge-covariant basis of one-forms on $\kflat_\text{col}$
is $(dx^\m,dy^\m,d\lambda_\a{}^I,D\psi^i)$,
where $D$ is the gauge-covariant exterior derivative.
The horizontal vector field $N \in \Gamma(\kflat_\text{col})$
is uniquely defined by 
the interior products
$\i_N : (dx^\m,dy^\m,d\lambda_\a{}^I,D\psi^i) \mapsto (y^\m,0,0,0)$.
It follows that $N$ is the generator of gauge-covariant translations \cite{gde}.

The Coulomb particle is described by
the color-sector symplectic potential
\eq{
	\label{stheta0}
	\stheta_{(0)}
	\mem=\,
		i\mem \bpsi_i\mem D\psi^i
	\,=\,
		i\mem \bpsi_i\mem d\psi^i
		+ q_a A^a
	\,,
}
where $q_a {\:=\:} i\mem \bpsi_i\mem (t_a)^i{}_j\hem \psi^j$
is the adjoint color charge.
A deviated Coulomb particle is described by
$\mathe^{\pounds_N} \stheta_{(0)}$:
\eq{
	\label{stheta0.dev}
		\cos(\pounds_N)\bigbig{
			i\mem \bpsi_i\mem D\psi^i
		}
		+ 
		\sinc(\pounds_N)\bigbig{
			q_a\mem \i_N F^a
		}
	\,.
}

In \eqref{stheta0.dev}, the $\cos(\pounds_N)$ term
implements two electric charges separated apart
at $x\pm iy$.
The $\sinc(\pounds_N)$, on the other hand,
inserts a thin electric flux tube between these electric charges.
We apply electric-magnetic duality
on the $\sinc(\pounds_N)$ term
via Hodge dual on the field strength:
\eq{
	\label{stheta1}
	\stheta_{(1)}
	\mem=\,
		\cos(\pounds_N)\bigbig{
			i\mem \bpsi_i\mem D\psi^i
		}
		+ 
		\sinc(\pounds_N)\bigbig{
			q_a\mem \i_N {*}F^a
		}
	\,.
	\kern-0.25em
}
As a result, we obtain the symplectic potential
describing two electric charges at $x\pm iy$
joined by a magnetic flux tube, i.e., a Dirac string
(cf. \rcite{GenSymGrav}).
Upon Gilbert-Amp\`ere duality, this describes a pair of self-dual and anti-self-dual dyons.
This pair is nothing other than 
the 
(dynamical version of)
{\Kerr} solution:
repeat the exercise in \rcite{nja}.

\begin{figure}[t]
	\centering
	\label{eq:Liecd-rKerr}
	\adjustbox{valign=c}{\begin{tikzpicture}
	    \node[empty] (O) at (0,0) {};
	    \node[empty] (X) at (3.75, 0) {};
	    \node[empty] (x) at (3.0, 0) {};
	    \node[empty] (Y) at (0, -0.95) {};
	    \node[w] (a00) at ($(O)$) {$i\mem \bpsi_i\hem D\psi^i$};
	    \node[w] (a01) at ($(O)+1*(x)$) {$0$};
	    \node[w] (a10) at ($(O)+1*(Y)$) {$q_a (\i_N F^a)$};
	    \node[w] (a11) at ($(O)+1*(Y)+1*(x)$) {$0$};
	    \node[w] (a20) at ($(O)+2*(Y)$) {$q_a (\i_ND\mem \i_N F^a)$};
	    \node[w] (a21) at ($(O)+2*(Y)+1*(x)$) {$0$};
	    \node[w] (a30) at ($(O)+3*(Y)$) {$\vdots$};
	    \node[w] (a2K) at ($(O)+2*(Y)-1*(X)$) {$q_a (\i_N {\ast F}^a)$};
	    \node[w] (a3K) at ($(O)+3*(Y)-1*(X)$) {$q_a (\i_ND\mem \i_N {\ast F^a})$};
	    \node[w] (a4K) at ($(O)+4*(Y)-1*(X)$) {$\vdots$};
	    \node[w] (a2k) at ($(O)+2*(Y)-1*(X)+(x)$) {$0$};
	    \node[w] (a3k) at ($(O)+3*(Y)-1*(X)+(x)$) {$0$};
	    \node[w] (phantom-a00) at ($(O)$) {$\phantom{\big|}$};
	    \node[w] (phantom-a01) at ($(O)+1*(x)$) {};
	    \node[w] (phantom-a10) at ($(O)+1*(Y)$) {};
	    \node[w] (phantom-a11) at ($(O)+1*(Y)+1*(x)$) {};
	    \node[w] (phantom-a20) at ($(O)+2*(Y)$) {};
	    \node[w] (phantom-a21) at ($(O)+2*(Y)+1*(x)$) {};
	    \node[w] (phantom-a30) at ($(O)+3*(Y)$) {};
	    \node[w] (phantom-a31) at ($(O)+3*(Y)+1*(x)$) {};
	    \node[w] (phantom-a2K) at ($(O)+2*(Y)-1*(X)$) {};
	    \node[w] (phantom-a3K) at ($(O)+3*(Y)-1*(X)$) {};
	    \node[w] (phantom-a4K) at ($(O)+4*(Y)-1*(X)$) {};
	    \node[w] (phantom-a2k) at ($(O)+2*(Y)-1*(X)+(x)$) {};
	    \node[w] (phantom-a3k) at ($(O)+3*(Y)-1*(X)+(x)$) {};
	    \draw[->] (a00)--(a01) node[midway,above] {\scriptsize \smash{${d}\mem\i_N$}\vphantom{d}};
	    \draw[->] (a10)--(a11) node[] {};
	    \draw[->] (a20)--(a21) node[] {};
	    \draw[->] (phantom-a00)--(phantom-a10) node[midway,left] {\scriptsize $\i_N{d}$};
	    \draw[->] (phantom-a10)--(phantom-a20) node[] {};
	    \draw[->] (phantom-a20)--(phantom-a30) node[] {};
	    \draw[<->] (phantom-a10)--(a2K) node[midway,above] {\adjustbox{raise=1.5pt}{\scriptsize dual}};
	    \draw[->] (phantom-a2K)--(phantom-a3K) node[] {};
	    \draw[->] (phantom-a3K)--(phantom-a4K) node[] {};
	    \draw[->] (a2K)--(a2k) node[] {};
	    \draw[->] (a3K)--(a3k) node[] {};
	\end{tikzpicture}}
	\caption{
		The ``$\i_N d / {*}$ sequence'' for {\Kerr}.
		A tree of one-forms
		emanates from
		the Coulomb particle's
		color-sector symplectic potential,
		$i\mem \bpsi_i\hem D\psi^i$.
	}
	\label{stree-of-life}
\end{figure}

\eqref{stheta1}
pinpoints a unique gauge theory interaction
to all orders in spin and gauge coupling.
From \fref{stree-of-life},
it can be seen that
\eqref{stheta1} can be represented as
\eq{
	\label{ssum1}
	\stheta_{(1)}
	\mem&=\,
	\stheta_{(0)}
	+\mem\hhem
		\sum_{\ell=1}^\infty
			\frac{1}{\ell!}\,
				q_a\mem
				\bigbig{\hnem
					(\i_N\hem D)^{\ell-1} \i_N\mem {*^\ell\hnem} F^a
				\hhnem}
	\,,
}
where ${*^\ell}$ means to act on the internal Hodge star $\ell$ times.
\eqref{ssum1} parallels \eqref{sum1}
by identifying spin angular momentum as ``the gravitational charge''
and Riemann curvature two-form as the ``gravitational field strength'':
gravity as a gauge theory of Lorentz group.

By methods in differential geometry \cite{gde},
\eqref{ssum1} evaluates to
\vspace{-1.0\baselineskip}
\begin{widetext}
\vspace{-1.0\baselineskip}
\vspace{-0.35\baselineskip}
\eq{
	\label{stheta-earth}
	\stheta_{(1)}
	\mem&=\,
	i\mem\bpsi_i D\psi^i
	+\mem\hhem
		\sum_{\ell=1}^\infty
			\frac{1}{\ell!}\,
				q_a
	    	 		\bb{\hnem
					\bigbig{\hnem
						{*}^\ell P_\ell
					\hnem}
					{}^a{}_\s\, dx^\s
					+
					(\ell\mminus1)\mem
					\bigbig{\hnem
						{*}^\ell P_{\ell-1}
					\hnem}
					{}^a{}_\s\, dy^\s
	    	 		}
	\,,
}
\vspace{-0.55\baselineskip}
\end{widetext}
\phantom{.}

\vspace{-2.1\baselineskip}\noindent
where
the so-called $P$-tensors \cite{gde} are defined as
\begin{align}
    \label{Ptensor}
    ({*}^\ell P_j)^a{}_\s
    \,=\,
	  {*}^\ell F^a{}_{\r_1\s;\r_2;\cdots;\r_j}\hnem(x)
	  \, y^{\r_1}{\cdots}y^{\r_j}
    \,.
\end{align}

\eqref{stheta-earth} provides the action of the {\Kerr} particle
coupled to generic nonabelian gauge field configurations (earth).
It is manifestly real and parity-symmetric
and is based on the conventional ``spacetime $+$ spin'' picture.

Newman-Janis shift manifests in heaven.
On the support of self-duality ${*}F^a = +i\mem F^a$,
\eqref{ssum1} becomes
\eqsplit{
	\label{rkerr.heaven}
	\stheta_{(1)}
	\mem&=\,
		\mathe^{i\pounds_N} \stheta_{(0)}
	\,=\,
		i\mem \tpsi_{i'} D\psi^{i\sprime}
	\,,
}
which localizes on $z^\m = x^\m + iy^\m$.
Here, we have complexified $\V$
(as a real vector space)
to $\V^\C$.

There is an obstruction in simultaneously manifesting
the Newman-Janis shifts in earth for both self-dual and anti-self-dual sectors.
This obstruction is due to the very nonlinearity of nonabelian gauge theory
and poses a massive analog of the googly problem.
Concretely, it is impossible to write the earthly {\Kerr} action
as a sum of two separate worldline actions localized respectively at
$z^\m$ and $\bz^\m$,
since
$P_\ell$ inevitably exhibits
nonlinear mixing between self-dual and anti-self-dual modes
from $\ell \geq 2$.

Note that there is no such googly problem in abelian gauge theory,
as the gauge potential cleanly splits into self-dual and anti-self-dual parts
and the {\Kerr} symplectic potential is linear in fields.

Still, we can manifest the spinspacetime localization
on either $z^\m$ or $\bz^\m$.
By mimicking \eqref{wsform}
in terms of $\bigbig{(\mathe^{i\pounds_N} \mminus 1)/\pounds_N}\mem q_a\hem (\i_N F^\pm{}^a)$,
it follows that
\vspace{-1.0\baselineskip}
\begin{widetext}
\vspace{-1.0\baselineskip}
\vspace{-0.25\baselineskip}
\eqsplit{
\label{rkerr.googly}
	\stheta_{(1)}
	\mem&=\,
	\mathe^{i\pounds_N}\mem
	\bb{
		i\mem \tpsi_i D\psi^i
		+ 
		\frac{
			\mathe^{-2i\pounds_N} \mminus 1
		}{\pounds_N}
		\,q_a\mem (\i_NF^-{}^a)
	}
	\,,\\
    	\mem&=\,
		i\mem \tpsi_{i'} D\psi^{i\sprime}
		+\mem\hhem
			\sum_{\ell=1}^\infty
				\frac{(-2i)^\ell}{\ell!}\,
					q_{a'}
		    	 		\bb{\hnem
						\bigbig{\hnem
							P^-_\ell
						\hnem}
						{}^{a\sprime}{}_\s\, dz^\s
						+
						(\ell\mminus1)\mem
						\bigbig{\hnem
							P^-_{\ell-1}
						\hnem}
						{}^{a\sprime}{}_\s\, dy^\s
		    	 		}
 	 \,,
}
\vspace{-0.85\baselineskip}
\end{widetext}
\phantom{.}

\vspace{-2.5\baselineskip}\noindent
where 
$F^\pm{}^a := \frac{1}{2}\mem (F^a {\,\mp\,} i\mem {*}F^a)$
and
$(P_\ell^\pm) := \minie\mem (P_\ell {\,\mp\,} i\mem {*} P_\ell)$.

\eqref{rkerr.googly} provides
the ``radical'' spinspacetime formulation of the {\Kerr} action,
i.e., a googly formulation.
It describes
\textit{both} self-dual and anti-self-dual couplings
as an action localized on the \textit{holomorphic} worldline $z^\m$.
It reduces to \eqref{rkerr.heaven} in the self-dual limit.
It treats self-dual and anti-self-dual modes,
and the $z^\m$ and $\bz^\m$ worldlines,
asymmetrically.
In perturbation theory, it manifests spin exponentiation of the positive-helicity (but not negative-helicity) Compton amplitudes.

We had computed the mixed-helicity Compton amplitudes in Yang-Mills theory and gravity,
the results of which will be published very soon.

\newpage
\paragraph{Acknowledgements}%
J.-H.K. is supported by the Department of Energy (Grant No.~DE-SC0011632) and by the Walter Burke Institute for Theoretical Physics.

\bibliography{references.bib}

\end{document}